\documentclass[%
 reprint,
 amsmath,amssymb,
 aps,
]{revtex4-2}

\usepackage{graphicx}
\usepackage{dcolumn}
\usepackage{bm}

\usepackage{color} 
\usepackage[dvipsnames, svgnames, x11names]{xcolor} 
\usepackage{enumerate} 

\begin{document}

\preprint{APS/123-QED}

\title{Asymmetric mode-pairing quantum key distribution}

\author{Zeyang Lu\textsuperscript{1}}

\author{Gang Wang\textsuperscript{1}}

\author{Chan Li\textsuperscript{1}}

\author{Zhu Cao\textsuperscript{1}}
\email{caozhu@ecust.edu.cn}

\affiliation{
\textup{\textsuperscript{1}}Key Laboratory of Smart Manufacturing in Energy Chemical Process,\\ Ministry of Education,\\ East China University of Science and Technology, Shanghai, 200237, China
}

\date{\today}

\begin{abstract}
Mode-pairing quantum key distribution (MP-QKD) can surpass the repeaterless rate-transmittance bound (Pirandola-Laurenza-Ottaviani-Banchi bound) without requiring global phase locking, exhibiting remarkable flexibility.
However, MP-QKD necessitates equal communication distances in two channels, which is a challenging requirement in practical applications.
To address this limitation, we extend the original MP-QKD to asymmetric cases.
Our decoy-state estimation confirms that asymmetric channel transmittances and asymmetric intensities do not compromise the security of the protocol. 
We focus on the pulse-intensity relationship, a key factor for optimizing the performance of asymmetric MP-QKD.
Unlike previous asymmetric protocols, the intensities of different bases in asymmetric MP-QKD cannot be decoupled.
We introduce an optimal-pulse-intensity method, adaptable to various scenarios, to enhance key rates by calculating ideal pulse intensities.
Simulation results in various representative scenarios indicate that our method effectively reduces the impact of asymmetric channel distances on MP-QKD performance, enhancing its practical applicability.
\end{abstract}

\maketitle

\section{Introduction}\label{sec1}
Quantum key distribution (QKD) is a quantum cryptography technology that enables two parties (commonly denoted as Alice and Bob) to generate a shared secret key known only to them, which can be used for the encryption and decryption of messages. 
It relies on the principles of quantum mechanics to guarantee the security of the key distribution process \cite{lo1999unconditional, shor2000simple}.
The introduction of the first QKD protocol, Bennett-Brassard 1984 (BB84) \cite{bennett2014quantum}, sparked a surge in QKD research \cite{ekert1991quantum, hwang2003quantum, wang2005beating, lo2005decoy}.
Indeed, several successful attacks have exploited security vulnerabilities in real-world devices to target QKD systems, shedding light on the limitations of the general QKD protocol \cite{qi2007time, makarov2006effects, xu2010experimental, zhao2008quantum, gerhardt2011full}.

To enhance the security of QKD, device-independent QKD (DI-QKD) \cite{acin2007device, pironio2009device} was proposed. 
This protocol reduces the assumptions needed for secure communication. 
However, DI-QKD and some of its improved protocols \cite{ho2020noisy, tan2020advantage, schwonnek2021device, xu2022device} place significant demands on devices.
In contrast, measurement-device-independent QKD (MDI-QKD) \cite{lo2012measurement} reduces the demand for detector efficiency. 
Several noteworthy experimental implementations were documented in Refs. \cite{da2013proof, liu2013experimental, tang2014experimental, tang2014measurement, yin2016measurement}.
Nonetheless, the efficiency of key generation is significantly impacted by the transmittance of the optical channel.
The asymptotic key rate is limited by the repeaterless rate-transmittance bound [Pirandola-Laurenza-Ottaviani-Banchi (PLOB) bound] \cite{pirandola2017fundamental}.
The PLOB bound sets a limit on how much information can be securely transmitted over a quantum channel. 
Breaking through this bound means that more information can be transmitted securely, enhancing the performance of quantum communication systems (e.g., the key rate).
This issue is addressed by the twin-field QKD (TF-QKD) proposal \cite{lucamarini2018overcoming}, which departs from previous coincidence measurements and, instead, leverages single-photon interference to surpass the PLOB bound.
Additionally, many variants of TF-QKD have been proposed, including phase-matching QKD \cite{ma2018phase, lin2018simple}, sending-or-not-sending QKD \cite{wang2018twin}, and no-phase postselection TF-QKD \cite{cui2019twin}. 
Related experiments, demonstrating the superior performance of TF-QKD and its variants, were proposed in Refs. \cite{minder2019experimental, liu2019experimental, pittaluga2021600, chen2021twin, yin2021twin}.
However, implementing TF-QKD and its variants necessitates the use of global phase locking, which significantly increases the need for experimental equipment and thus reduces the practicality of the protocol.

Recently, mode-pairing QKD (MP-QKD) \cite{zeng2022mode} and its experiment \cite{zhu2023experimental} were proposed, incorporating several enhancements over TF-QKD.
On the one hand, MP-QKD surpasses the PLOB bound by encoding key information using relative phases, thereby obviating the need for global phase locking.
On the other hand, it offers the flexibility to switch between different pairing schemes.
As a result, MP-QKD not only improves protocol performance but also enhances the practicality and flexibility of quantum communication.
However, similar to most protocols, MP-QKD needs to be implemented symmetrically.
This entails ensuring that the intermediate party is equidistant from both sides of the communication.
In real-life situations, achieving the condition of equal distance is often challenging. 
Moreover, the variation in distances between the parties results in differing channel transmittances.
If the original protocol had been employed, this would have significantly reduced the key rate.

In this work, we extend the original MP-QKD to accommodate asymmetric scenarios and explore how to get better performance in such cases.
Through the decoy-state estimation, we confirm that the security of the asymmetric MP-QKD is not affected by the asymmetric channel transmittances and asymmetric intensities.
The practical way to improve the performance of asymmetric protocols is to select the appropriate pulse intensities.
Unlike previous protocols that preselect pulses for either key generation (typically referred to as the $Z$ basis) or decoy-state estimation (typically referred to as the $X$ basis), the original MP-QKD generates these two bases only after they have been measured.
In this context, previous asymmetric protocols \cite{wang2019asymmetric, grasselli2019asymmetric, wang2020simple, zhou2019asymmetric} can separate the pulse intensities across different bases.
In asymmetric MP-QKD, however, such decoupling of pulse intensities is unattainable, resulting in a consistent relationship between pulse intensities for both $Z$ and $X$ bases.
Hence, we develop an optimal-pulse-intensity method for asymmetric MP-QKD.
This method enhances the protocol performance by identifying pulse intensities that optimize the key rate.
We study the relationships of optimal intensities at various communication distances.
Since the variation of the maximal pairing interval affects the performance of MP-QKD, we investigate the trend of the optimal pulse intensities in response to this variation.
Furthermore, we plot the optimal pulse intensities to verify their dependence on various factors.
A straightforward approach to address asymmetric protocols involves adding extra fiber to the closer side, thereby maintaining equal transmittance on both sides. 
However, this adjustment will result in a notably reduced key rate due to the overall lower transmittance.
We compare and analyze this method alongside the optimal-pulse-intensity approach for various distance differences.
Additionally, we simulate the performance of asymmetric MP-QKD using the optimal-pulse-intensity method for different pairing intervals.
Finally, we show this protocol's tolerance for misalignment errors across different differences between two distances.
These simulations are conducted in the asymptotic case to demonstrate the performance of asymmetric MP-QKD effectively.

The structure of this paper is summarized as follows.
In Sec. \ref{sec2}, we describe the operational steps of the asymmetric MP-QKD and present its schematic diagram.
In Sec. \ref{sec3}, we demonstrate the security of asymmetric MP-QKD by employing decoy-state estimation.
In Sec. \ref{sec4}, we discuss the method to improve the performance of the protocol by selecting the optimal pulse intensities and analyzing the optimal intensities for different scenarios.
In Sec. \ref{sec5}, we simulate asymmetric MP-QKD in various scenarios.
In Sec. \ref{sec6}, we present our conclusions and outlook.

\section{Asymmetric Mode-pairing quantum key distribution} \label{sec2}

\begin{figure*}[htbp] 
\centering
\includegraphics[scale=0.235]{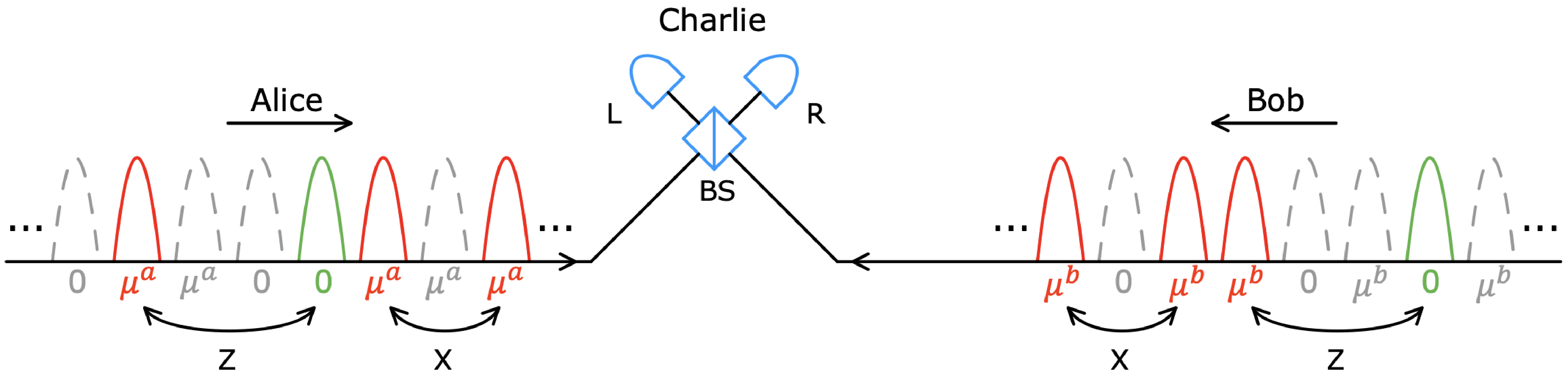}
\caption{An example setup for asymmetric MP-QKD. 
The communicating parties, Alice and Bob, randomly and uniformly prepare weakly coherent pulses with varying intensity ($\mu^{a(b)}_{i}\in \{0,\mu^{a(b)}\}$) and phase ($\phi^{a(b)}_{i}\in[0,2\pi)$), which they then send to the intermediate party, Charlie. 
After Charlie performs an interference measurement through the beam splitter (BS) and announces the outcomes for detectors L and R, Alice and Bob proceed to pair the successfully detected pulses and determine their coding bases according to specific rules. 
$Z$ bases are used for key generation, while other data are used for parameter estimation.
The distances from Alice to Charlie and from Bob to Charlie are denoted as $L_{a}$ and $L_{b}$, respectively. 
Without loss of generality, it is assumed that $L_{a}<L_{b}$.
For clear differentiation, the intensity value of each pulse is labeled beneath its corresponding pulse.
In addition, pulses successfully detected are denoted by solid lines, while undetected pulses are indicated by dashed lines.
}
\label{fig_protocol}
\end{figure*}

The schematic of the asymmetric MP-QKD setup is shown in Fig. \ref{fig_protocol}.
The details of this protocol are presented as follows.

\text{(1)} \textit{Preparation.} At each time bin $i\in \{1,2,...,N\}$, Alice (Bob) prepares a weak coherent pulse $|\sqrt{\mu^{a}_{i}}e^{i\phi ^{a}_{i}}\rangle$ $\left ( |\sqrt{\mu^{b}_{i}}e^{i\phi ^{b}_{i}}\rangle \right )$, where the intensity $\mu^{a}_{i}$ $\left ( \mu^{b}_{i} \right )$ and the phase $\phi^{a}_{i}$ $\left ( \phi^{b}_{i} \right )$ are randomly and uniformly selected from $\{ 0,\mu^{a} \}$ $\left ( \{ 0,\mu^{b} \}\right )$ and $[0,2\pi)$, respectively.

\text{(2)} \textit{Measurement.} For each time bin $i$, Alice and Bob send their pulses to an untrusted node named Charlie, which is located between them.
The communication distances from Alice to Charlie and from Bob to Charlie are denoted as $L_{a}$ and $L_{b}$, respectively. 
Without loss of generality, it is assumed that $L_{a}<L_{b}$.
The corresponding channel transmittances are $\eta_{a}$ and $\eta_{b}$, respectively. 
Additionally, Charlie performs single-photon interference measurements on the two received pulses and publicly announces the measurement outcomes for detectors L and R.

\text{(3)} \textit{Mode pairing.} Alice and Bob repeat the first two steps for $N$ rounds. 
In rounds where successful detection occurs, with only one of the two detectors clicking, Alice and Bob group every two of these detected rounds into pairs. 
Note that the number of pulses between the two paired rounds should not exceed the maximal pairing interval $\lambda$.

\text{(4)} \textit{Basis sifting.} For two paired rounds indexed as $i$ and $j$, Alice (Bob) labels them as a $Z$ pair if their intensities satisfy $\mu^{a}_{i}+\mu^{a}_{j}=\mu^{a}$ $(\mu^{b}_{i}+\mu^{b}_{j}=\mu^{b})$, as a $X$ pair if their intensities satisfy $\mu^{a}_{i}=\mu^{a}_{j}=\mu^{a}$ $(\mu^{b}_{i}=\mu^{b}_{j}=\mu^{b})$, or as “0” pair if their intensities satisfy $\mu^{a}_{i}=\mu^{a}_{j}=0$ $(\mu^{b}_{i}=\mu^{b}_{j}=0)$. 
Alice and Bob announce their respective bases of each successful pair. 
If the announced bases are either all $X$ or all $Z$ and have the same time bins, they are retained; otherwise, they are discarded.

\text{(5)} \textit{Key mapping.} For every $Z$ pair located at positions $i$ and $j$, Alice (Bob) sets the raw key bit as $\kappa^{a}=0$ $\left ( \kappa^{b}=1 \right )$ if $(\mu^{a}_{i},\mu^{a}_{j})=(0,\mu^{a})$ $\left [ (\mu^{b}_{i},\mu^{b}_{j})=(0,\mu^{b}) \right ]$ and as $\kappa^{a}=1$ $\left ( \kappa^{b}=0 \right )$ if $(\mu^{a}_{i},\mu^{a}_{j})=(\mu^{a},0)$ $\left [ (\mu^{b}_{i},\mu^{b}_{j})=(\mu^{b},0) \right ]$. 
For every $X$ pair located at positions $i$ and $j$, Alice obtains the raw key bit from the relative phase $(\phi^{a}_{j}-\phi^{a}_{i})=\theta^{a}+\pi\kappa^{a}$, where the key is $\kappa^{a}=\left \lfloor \left [ (\phi^{a}_{j}-\phi^{a}_{i})/\pi \right ]\text{mod}2 \right \rfloor$ and the alignment angle is $\theta^{a}=(\phi^{a}_{j}-\phi^{a}_{i})\text{mod}\pi$. 
Bob extracts the raw key bit $\kappa^{b}$ and computes $\theta^{b}$ in a similar manner. 
If the detector click pattern for the $X$ pair is either $(L,L)$ or $(R,R)$, Bob retains the key $\kappa^{b}$. 
However, if the click pattern is either $(L,R)$ or $(R,L)$, Bob flips $\kappa^{b}$. 
Furthermore, Alice and Bob announce the alignment angles $\theta^{a}$ and $\theta^{b}$ in the $X$ pairs. 
If $\theta^{a}=\theta^{b}$, they keep the paired rounds; otherwise, they discard the paired rounds.

\text{(6)} \textit{Parameter estimation.} Alice and Bob use the $Z$ pairs to generate the raw key. 
They estimate the expected single-photon pair ratio $\bar{q}_{(1,1)}$ in all $Z$ pairs, the phase error $e_{(1,1)}$ of the single-photon pairs using decoy-state methods, and the quantum bit error rate $e^{(\mu^{a},\mu^{b}),Z}$ of the $Z$ pairs.

\text{(7)} \textit{Postprocessing.} After applying error correction and privacy amplification to the raw key bits, Alice and Bob obtain the final secret key.

\section{Decoy state estimation} \label{sec3}
In this section, we show that the security of asymmetric MP-QKD is not compromised by asymmetric channel transmittances and asymmetric intensities.
The security proof for the original MP-QKD is provided in Ref. \cite{zeng2022mode}.
On this basis, to analyze the security of asymmetric MP-QKD, it is necessary to estimate both the bit error rate and the phase error rate.
The bit error rate can be directly obtained from the experiment, while the phase error rate cannot be directly measured.
Therefore, we estimate the phase error rate of asymmetric MP-QKD by extending the decoy-state estimation from the original MP-QKD to the asymmetric case.
Note that the decoy-state estimation is analyzed with the infinite key size.

We employ decoy-state analysis with three different pulse intensities \cite{zhang2017improved}.
For each time bin $i$, Alice (Bob) randomly selects the pulse intensities $\mu^{a}_{i}$ ($\mu^{b}_{i}$) from the set $\{0,\nu^{a},\mu^{a}\}$ ($\{0,\nu^{b},\mu^{b}\}$) with corresponding probabilities $s_{0}$, $s_{\nu^{a}}$ ($s_{\nu^{b}}$), and $s_{\mu^{a}}$ ($s_{\mu^{b}}$), where the sum of these probabilities equals 1.
To simplify the discussion, the probabilities of the two parties choosing an intensity are set to $s_{\nu^{a}}=s_{\nu^{b}}$ and $s_{\mu^{a}}=s_{\mu^{b}}$.

Suppose Alice and Bob each send $N$ pulses, with $N$ being sufficiently large. 
Based on these pulses, Alice and Bob pair two locations indexed as $i$ and $j$, including the locations with unsuccessful clicks.
The intensity vector of the $(i,j)$ pair is denoted as
\begin{equation}
    \begin{split}
        \vec{\mu}=(\mu^{a}_{i}+\mu^{a}_{j},\mu^{b}_{i}+\mu^{b}_{j}),
    \end{split}
\end{equation}
where $\mu^{a(b)}_{i},\mu^{a(b)}_{j}\in\{0,\nu^{a(b)},\mu^{a(b)}\}$ and, consequently, $\mu^{a(b)}_{i}+\mu^{a(b)}_{j}\in\{0,\nu^{a(b)},\mu^{a(b)},2\nu^{a(b)},\nu^{a(b)}+\mu^{a(b)},2\mu^{a(b)}\}$.
The probability of Alice and Bob sending intensities $\vec{\mu}$ for the $(i,j)$ pair is denoted as
\begin{equation}
    \begin{split}
        q^{\vec{\mu}} =\sum_{(\mu^{a}_{i}+\mu^{a}_{j},\mu^{b}_{i}+\mu^{b}_{j})=\vec{\mu}}s_{\mu^{a}_{i}}s_{\mu^{a}_{j}}s_{\mu^{b}_{i}}s_{\mu^{b}_{j}}.
    \end{split}
\end{equation} 
This probability is independent of the measurement results announced by Charlie.

Alice and Bob can carry out photon-number measurements on the ancillary systems.
For each pair of locations ($i,j$), the results of the photon-number measurements performed by Alice and Bob are denoted as $k^{a}$ and $k^{b}$, respectively.
We denote the photon numbers on the ($i,j$) pair as $\vec{k}=(k^{a},k^{b})$.
Provided that Alice and Bob send intensities $\vec{\mu}$ for the ($i,j$) pair, the probability of their photon-number measurements yielding $\vec{k}$ is denoted as
\begin{equation} \label{eq_Pr_kmu}
    \begin{split}
        \text{Pr}(\vec{k}|\vec{\mu})=e^{-(\mu^{a}_{i}+\mu^{a}_{j})-(\mu^{b}_{i}+\mu^{b}_{j})} \frac{(\mu^{a}_{i}+\mu^{a}_{j})^{k^{a}}(\mu^{b}_{i}+\mu^{b}_{j})^{k^{b}}}{k^{a}!k^{b}!},
    \end{split}
\end{equation}
which consists of the product of two Poisson distributions, since the intensity settings of the two parties are independent.

The probability of Alice and Bob measuring the photon number on the ($i,j$) pair and obtaining the result $\vec{k}$ is denoted as $q_{\vec{k}}$.
Given the result $\vec{k}$, the probability for the intensity setting on the $(i,j)$ pair to be $\vec{\mu}$ is expressed as
\begin{equation} \label{eq_Pr_muk}
    \begin{split}
        \text{Pr}(\vec{\mu}|\vec{k}) = \frac{q^{\vec{\mu}}\text{Pr}(\vec{k}|\vec{\mu})}{q_{\vec{k}}} = \frac{q^{\vec{\mu}}\text{Pr}(\vec{k}|\vec{\mu})}{\sum_{\vec{\mu}'}q^{\vec{\mu}'}\text{Pr}(\vec{k}|\vec{\mu}')},
    \end{split}
\end{equation}
where the values of $\vec{\mu}'$ are taken to be the same as those of $\vec{\mu}$. The distinct notation $\vec{\mu}'$ is used to avoid ambiguity in the summation.
$\text{Pr}(\vec{\mu}|\vec{k})$ and $\text{Pr}(\vec{k}|\vec{\mu})$, just like $q^{\vec{\mu}}$, are the prior probability distributions and are independent of Charlie's measurement outcomes.

After Charlie completes measurements and announces the results for detectors L and R, Alice and Bob perform mode pairing and basis sifting according to the pairing strategy.
If the intensity vector $\vec{\mu}$ on the $(i,j)$ pair satisfies
\begin{equation}
    \begin{split}
        \mu^{a}_{i}\mu^{a}_{j} =\mu^{b}_{i}\mu^{b}_{j}=0,\ \mu^{a}_{i}+\mu^{a}_{j}+\mu^{b}_{i}+\mu^{b}_{j}\neq 0,
    \end{split}
\end{equation}
then it is a $Z$ pair.
We concentrate on decoy-state estimation in the $Z$ pair.
Estimation in the $X$ pair can be derived using a similar approach.

The decoy-state estimation discussed next is performed on $Z$ pairs.
Suppose Alice and Bob get $M^{Z}$ rounds of $Z$ pairs with successful detection, and among these, $E^{Z}$ rounds are erroneous.
Moreover, we denote the total number of pairs with intensity setting $\vec{\mu}$ as $M^{\vec{\mu},Z}$ and the corresponding number of erroneous pairs as $E^{\vec{\mu},Z}$.
The total and erroneous numbers of pairs with photon numbers $\vec{k}$ are denoted as $M^{Z}_{\vec{k}}$ and $E^{Z}_{\vec{k}}$, respectively. 
Among these pairs, $M^{\vec{\mu},Z}_{\vec{k}}$ and $E^{\vec{\mu},Z}_{\vec{k}}$ denote the pairs with intensity setting $\vec{\mu}$.
These values satisfy
\begin{equation}
    \begin{split}
        M^{Z}&=\sum_{\vec{\mu}}M^{\vec{\mu},Z} =\sum_{\vec{k}}M^{Z}_{\vec{k}} =\sum_{\vec{\mu}}\sum_{\vec{k}}M^{\vec{\mu},Z}_{\vec{k}},\\
        E^{Z}&=\sum_{\vec{\mu}}E^{\vec{\mu},Z} =\sum_{\vec{k}}E^{Z}_{\vec{k}} =\sum_{\vec{\mu}}\sum_{\vec{k}}E^{\vec{\mu},Z}_{\vec{k}}.
    \end{split}
\end{equation}
Throughout the protocol, Alice and Bob are aware of the values $M^{\vec{\mu},Z}$ and $E^{\vec{\mu},Z}$, but they remain unaware of the values $M^{Z}_{\vec{k}}$ and $E^{Z}_{\vec{k}}$, which are fixed after Charlie announces the results.

In practice, Alice and Bob first perform photon-number measurement, resulting in the outcome $\vec{k}$. 
Subsequently, both parties randomly choose the intensity setting $\vec{\mu}$ based on $\vec{k}$. 
Therefore, the intensity setting $\vec{\mu}$ is solely dependent on $\vec{k}$ and is independent of the result that Charlie announces.
In response, when considering all the generated $Z$ pairs where the photon-number measurement results in $\vec{k}$, the expected ratio of different intensity settings should be the same as the ratio of emitted states, i.e., 
\begin{equation}
    \begin{split}
        \frac{M^{\vec{\mu},Z}_{\vec{k}}}{M^{\vec{\mu}',Z}_{\vec{k}}} = \frac{\text{Pr}(\vec{\mu},\vec{k})}{\text{Pr}(\vec{\mu}',\vec{k})} = \frac{q^{\vec{\mu}}\text{Pr}(\vec{k}|\vec{\mu})}{q^{\vec{\mu}'}\text{Pr}(\vec{k}|\vec{\mu}')}.
    \end{split}
\end{equation}

Within the set of $M^{Z}_{\vec{k}}$ pairs with the photon number of $\vec{k}$, the number of pairs with the intensity setting $\vec{\mu}$ is denoted as the random variable $\mathcal{M}^{\vec{\mu},Z}_{\vec{k}}$, which is determined by the ancillary systems of Alice and Bob.
Based on the preceding analysis, the expected ratio of the intensity setting $\vec{\mu}$ is expressed as
\begin{equation} \label{eq8}
    \begin{split}
        \mathbb{E}\left( \frac{\mathcal{M}^{\vec{\mu},Z}_{\vec{k}}}{M^{Z}_{\vec{k}}} \right) = \text{Pr}(\vec{\mu}|\vec{k}) = \frac{q^{\vec{\mu}}\text{Pr}(\vec{k}|\vec{\mu})}{\sum_{\vec{\mu}'}q^{\vec{\mu}'}\text{Pr}(\vec{k}|\vec{\mu}')},
    \end{split}
\end{equation}
where the variable $\mathcal{M}^{\vec{\mu},Z}_{\vec{k}}$ is used to characterize the intensity settings chosen by Alice and Bob.
Correspondingly, among the $E^{Z}_{\vec{k}}$ pairs with the photon number of $\vec{k}$, $\mathcal{E}^{\vec{\mu},Z}_{\vec{k}}$ denotes the number of pairs with the intensity setting $\vec{\mu}$.
The corresponding ratio is 
\begin{equation} \label{eq9}
    \begin{split}
        \mathbb{E}\left( \frac{\mathcal{E}^{\vec{\mu},Z}_{\vec{k}}}{E^{Z}_{\vec{k}}} \right) = \text{Pr}(\vec{\mu}|\vec{k}) = \frac{q^{\vec{\mu}}\text{Pr}(\vec{k}|\vec{\mu})}{\sum_{\vec{\mu}'}q^{\vec{\mu}'}\text{Pr}(\vec{k}|\vec{\mu}')}.
    \end{split}
\end{equation}
Based on Eqs. (\ref{eq8}) and (\ref{eq9}), we derive the following results:
\begin{equation} \label{eq_Emuk}
    \begin{split}
        \mathbb{E}[\mathcal{M}^{\vec{\mu},Z}_{\vec{k}}] &= \text{Pr}(\vec{\mu}|\vec{k})M^{Z}_{\vec{k}}, \\ \mathbb{E}[\mathcal{E}^{\vec{\mu},Z}_{\vec{k}}] &= \text{Pr}(\vec{\mu}|\vec{k})E^{Z}_{\vec{k}}.
    \end{split}
\end{equation}
The total and erroneous numbers of pairs with the intensity settings $\vec{\mu}$ are denoted as $\mathcal{M}^{\vec{\mu},Z}$ and $\mathcal{E}^{\vec{\mu},Z}$, respectively, where
\begin{equation} \label{eq_mu}
    \begin{split}
        \mathcal{M}^{\vec{\mu},Z} &= \sum_{\vec{k}}\mathcal{M}^{\vec{\mu},Z}_{\vec{k}}, \\ \mathcal{E}^{\vec{\mu},Z} &= \sum_{\vec{k}}\mathcal{E}^{\vec{\mu},Z}_{\vec{k}}.
    \end{split}
\end{equation}
Based on Eqs. (\ref{eq_Emuk}) and (\ref{eq_mu}), we can derive 
\begin{equation} \label{eq_Emu}
    \begin{split}
        \mathbb{E}[\mathcal{M}^{\vec{\mu},Z}] &= \sum_{\vec{k}}\text{Pr}(\vec{\mu}|\vec{k})M^{Z}_{\vec{k}}, \\ \mathbb{E}[\mathcal{E}^{\vec{\mu},Z}] &= \sum_{\vec{k}}\text{Pr}(\vec{\mu}|\vec{k})E^{Z}_{\vec{k}}.
    \end{split}
\end{equation}

For $Z$ pairs, the total and erroneous ratios of pairs with intensity setting $\vec{\mu}$ and photon numbers $\vec{k}$ are defined as
\begin{equation} \label{eq_r}
    \begin{split}
        &(m')^{\vec{\mu},Z}=\frac{\mathcal{M}^{\vec{\mu},Z}}{N^{\vec{\mu}}},\quad (e')^{\vec{\mu},Z}=\frac{\mathcal{E}^{\vec{\mu},Z}}{N^{\vec{\mu}}}, \\ &m^{Z}_{\vec{k}}=\frac{M^{Z}_{\vec{k}}}{N_{_{\vec{k}}}},\quad e^{Z}_{\vec{k}}=\frac{E^{Z}_{\vec{k}}}{N_{_{\vec{k}}}}, \\ &(m')^{\vec{\mu},Z}_{\vec{k}}=\frac{\mathcal{M}^{\vec{\mu},Z}_{\vec{k}}}{N^{\vec{\mu}}},\quad (e')^{\vec{\mu},Z}_{\vec{k}}=\frac{\mathcal{E}^{\vec{\mu},Z}_{\vec{k}}}{N^{\vec{\mu}}},
    \end{split}
\end{equation}
where $m(e)$ and $m'(e')$ are used to distinguish between different variables, $N^{\vec{\mu}}:=q^{\vec{\mu}}N$ is the number of rounds with intensity setting $\vec{\mu}$, and $N_{_{\vec{k}}}:=\sum_{\vec{\mu}}\text{Pr}(\vec{k}|\vec{\mu})N^{\vec{\mu}}$ is the number of rounds with photon numbers $\vec{k}$. 

Based on the analysis presented above, it can be concluded that
\begin{equation} \label{eq_E_rmu}
    \begin{split}
        \mathbb{E}[(m')^{\vec{\mu},Z}]=&\mathbb{E}[\frac{\mathcal{M}^{\vec{\mu},Z}}{N^{\vec{\mu}}}]=\frac{\mathbb{E}[\mathcal{M}^{\vec{\mu},Z}]}{N^{\vec{\mu}}}
        \\=&\frac{\sum_{\vec{k}}\text{Pr}(\vec{\mu}|\vec{k})M^{Z}_{\vec{k}}}{q^{\vec{\mu}}N}
        \\=&\sum_{\vec{k}}\frac{q^{\vec{\mu}}\text{Pr}(\vec{k}|\vec{\mu})}{\sum_{\vec{\mu}'}q^{\vec{\mu}'}\text{Pr}(\vec{k}|\vec{\mu}')}\frac{M^{Z}_{\vec{k}}}{q^{\vec{\mu}}N} 
        \\=&\sum_{\vec{k}}\text{Pr}(\vec{k}|\vec{\mu})\frac{M^{Z}_{\vec{k}}}{\sum_{\vec{\mu}'}\text{Pr}(\vec{k}|\vec{\mu}')(q^{\vec{\mu}'}N)} 
        \\=&\sum_{\vec{k}}\text{Pr}(\vec{k}|\vec{\mu})m^{Z}_{\vec{k}}, 
        \\\mathbb{E}[(e')^{\vec{\mu},Z}]=&\sum_{\vec{k}}\text{Pr}(\vec{k}|\vec{\mu})e^{Z}_{\vec{k}}.
    \end{split}
\end{equation}
Similarly, we can obtain
\begin{equation} \label{eq_E_rmuk}
    \begin{split}
        \mathbb{E}[(m')^{\vec{\mu},Z}_{\vec{k}}]=&\mathbb{E}[\frac{\mathcal{M}^{\vec{\mu},Z}_{\vec{k}}}{N^{\vec{\mu}}}]=\frac{\text{Pr}(\vec{\mu}|\vec{k})M^{Z}_{\vec{k}}}{q^{\vec{\mu}}N}
        \\=&\text{Pr}(\vec{k}|\vec{\mu})m^{Z}_{\vec{k}},
        \\\mathbb{E}[(e')^{\vec{\mu},Z}_{k}]=&\text{Pr}(\vec{k}|\vec{\mu})e^{Z}_{\vec{k}},
    \end{split}
\end{equation}
where $\text{Pr}(\vec{k}|\vec{\mu})$ is given by Eq. (\ref{eq_Pr_kmu}).
Note that the variation in $N$ does not affect Eqs. (\ref{eq_E_rmu}) and (\ref{eq_E_rmuk}).

$\mathbb{E}[(m')^{\vec{\mu},Z}]$ and $\mathbb{E}[(e')^{\vec{\mu},Z}]$ are obtained from the experiments.
Based on Eq. (\ref{eq_E_rmu}), one can estimate the lower bound of $m^{Z}_{(1,1)}$ and the upper bound of $e^{Z}_{(1,1)}$, represented as $m^{Z,L}_{(1,1)}$ and $e^{Z,U}_{(1,1)}$, respectively.
Subsequently, one can apply Eq. (\ref{eq_E_rmuk}) to derive the lower bound of $m^{\vec{\mu},Z}_{(1,1)}$ and the upper bound of $e^{\vec{\mu},Z}_{(1,1)}$, denoted as $m^{\vec{\mu},Z,L}_{(1,1)}$ and $e^{\vec{\mu},Z,U}_{(1,1)}$, respectively.

The above is the decoy-state analysis in the $Z$ basis.
Similar steps can be applied to obtain the bounds on the total and erroneous ratios of single-photon pairs in the $X$ basis ($m^{X,L}_{(1,1)}$ and $e^{X,U}_{(1,1)}$).
In decoy-state estimation, the key-rate formula for the asymptotic case in the asymmetric mode-pairing scheme is expressed as
\begin{equation} \label{dec_R}
    \begin{split}
        R = &m^{(\mu^{a},\mu^{b}),Z,L}_{(1,1)}\left[1-H\left(\frac{e^{X,U}_{(1,1)}}{m^{X,L}_{(1,1)}}\right) \right] \\&-fm^{(\mu^{a},\mu^{b}),Z}H(e^{(\mu^{a},\mu^{b}),Z}) \\= &m^{(\mu^{a},\mu^{b}),Z} \left \{ \bar{q}_{(1,1)} \left[1-H(e_{(1,1)}) \right] -fH(e^{(\mu^{a},\mu^{b}),Z}) \right \},
    \end{split}
\end{equation}
where $m^{(\mu^{a},\mu^{b}),Z,L}_{(1,1)}$ denotes the lower bound on the ratio of single-photon pairs with an intensity of $\vec{\mu}=(\mu^{a},\mu^{b})$ when Alice and Bob emit data in the $Z$ basis, $f$ denotes the error-correction efficiency, and $H$ is the binary entropy function.
The ratio of pairs with intensity $\vec{\mu}=(\mu^{a},\mu^{b})$ in the $Z$ basis, denoted as $m^{(\mu^{a},\mu^{b}),Z}$, and the quantum bit error rate, represented as $e^{(\mu^{a},\mu^{b}),Z}$, can be directly determined from experimental results.
The lower bound for the single-photon pair ratio in all $Z$ pairs $\bar{q}_{(1,1)}$ can be denoted as 
\begin{equation}
    \begin{split}
       \bar{q}_{(1,1)} = \frac{m^{(\mu^{a},\mu^{b}),Z,L}_{(1,1)}}{m^{(\mu^{a},\mu^{b}),Z}}.
    \end{split}
\end{equation}
$e_{(1,1)}$ denotes the upper bound on the phase error rate of single-photon pairs with the intensity of $\vec{\mu}=(\mu^{a},\mu^{b})$. 
It can be estimated directly in the asymptotic case from the following expression:
\begin{equation}
    \begin{split}
        e_{(1,1)} = \frac{e^{X,U}_{(1,1)}}{m^{X,L}_{(1,1)}}.
    \end{split}
\end{equation}

\section{Optimal-pulse-intensity method} \label{sec4}
In this section, we explore how to improve the performance of asymmetric MP-QKD by adjusting the pulse intensities ($\mu^{a}$ and $\mu^{b}$).
We develop a method that can calculate the optimal pulse intensities to maximize the key rate.
Moreover, we employ this calculation method to analyze the impact of channel transmittances ($\eta^{a}$ and $\eta^{b}$) and maximal pulse interval $\lambda$ on the optimal intensities.

In the original MDI-QKD and TF-QKD protocols \cite{lo2012measurement, lucamarini2018overcoming}, Alice and Bob preselect the $Z$ basis and $X$ basis randomly according to probabilities. 
However, in the original MP-QKD protocols, these two bases are determined after Charlie announces the measurement outcomes. 
Unlike the former scenario in which the pulse intensities in different bases could be decoupled, the intensities of the $Z$ and $X$ bases in MP-QKD are selected from the same set.
In asymmetric MDI-QKD and TF-QKD \cite{wang2019asymmetric, grasselli2019asymmetric, wang2020simple}, the intensities in the different bases have distinct impacts on specific parameters within the final key-rate formula.
Therefore, the correlation between the intensities in the respective bases varies when considering the maximal key rate.
In contrast, in the case of asymmetric MP-QKD, as the intensities of the $Z$ and $X$ bases are coupled, they all influence various parameters in the key-rate equation in the same way.
In this regard, instead of distinguishing between the $Z$ and $X$ cases, we can directly analyze the effect of their asymmetric intensities on the final key rate.

\subsection{Calculation method} \label{subsec1}
The key-rate simulation formula for the asymptotic case in asymmetric MP-QKD is expressed as
\begin{equation} \label{eq18}
    \begin{split}
        R=r_{p}(p,\lambda) r_{s} \left \{ \bar{q}_{(1,1)} \left [ 1-H(e_{(1,1)})\right ]-fH(e^{(\mu^{a},\mu^{b}),Z}) \right \},
    \end{split}
\end{equation}
where $r_{p}(p,\lambda)$ denotes the pairing rate in each round, $p$ denotes the probability of successful detection in each round, $\lambda$ is the maximal pairing interval, $r_{s}$ denotes the probability that two paired rounds are $Z$ pairs, $\bar{q}_{(1,1)}$ is the single-photon pair ratio in all $Z$ pairs, $H$ is the binary entropy function, $e_{(1,1)}$ is the phase error rate of the single-photon pairs, $f$ denotes the error-correction efficiency, and $e^{Z}_{(\mu^{a},\mu^{b})}$ is the quantum bit error rate of the $Z$ pairs.
Detailed expressions for these parameters are shown in Appendix \ref{appA}.

We investigate which values of two pulse intensities ($\mu^{a}$ and $\mu^{b}$) make the key rate optimal in scenarios involving channel transmittances ($\eta^{a}$ and $\eta^{b}$) and the maximal pairing interval $\lambda$.
These values can be determined through the following steps.

(1) Set both the ratio of channel transmittances $\eta^{a}/\eta^{b}$ and the communication distance from Alice to Charlie $L_{a}$ as constants (e.g., $\eta^{a}/\eta^{b}=\delta$ and $L_{a}>0$).
Without loss of generality, we focus on the case where Charlie is closer to Alice than to Bob ($L_{a}\leq L_{b}$), such that the range of the ratio is $\delta\geq 1$. 
In addition, set the maximal pairing interval $\lambda$ to be a constant, and its actual value range is $\lambda\geq 1$.

(2) Based on the parameters listed in Table \ref{tab1}, solve the following problem:
\begin{equation} \label{eq19}
    \begin{split}
        \max\quad &R\\\text{such that}\quad &0< \mu^{a}\leq 1,\\&0< \mu^{b}\leq 1,\\&0< R\leq 1,\\&L_{a}>0,\\&\frac{\eta^{a}}{\eta^{b}}=\delta\geq 1,\\&\lambda\geq 1,
    \end{split}
\end{equation} 
where $\mu^{a}$ and $\mu^{b}$ are variables representing the intensity of the weak coherent pulses prepared by Alice and Bob, respectively. 
Here, $0< R\leq 1$ is set to prevent invalid values in calculations.

(3) Record the values of $\mu^{a}$ and $\mu^{b}$, which maximize the key rate, and denote them as $\mu^{a}_{m}$ and $\mu^{b}_{m}$, respectively. 

The essence of the above problem lies in finding the extreme value of a multivariate function, which can be directly calculated using the corresponding methods.
Note that when $L_{a}=L_{b}$, the protocol falls under the symmetric case, wherein the optimal-pulse-intensity method remains applicable.

\begin{table}[htbp]
\caption{Parameters for performance analysis adopted from Ref. \cite{zeng2022mode}. 
$\eta_{d}$ denotes the detection efficiency, $\alpha$ denotes the attenuation coefficient of the fiber, $p_{d}$ denotes the dark count rate, $f$ denotes the error-correction efficiency, and $e_{d}$ denotes the misalignment error.}
\label{tab1}
\begin{tabular}{p{1.2cm}<{\centering} p{1.2cm}<{\centering} p{1.5cm}<{\centering} p{1.2cm}<{\centering} p{1.2cm}<{\centering}}
\hline \hline
$\eta_{d}$ & $\alpha$ & $p_{d}$ & $f$ & $e_{d}$ \\
\hline 
\noalign{\vskip 1pt} 
$20\%$ & $0.2$ & $1.2\times10^{-8}$ & 1.15 & $4\%$ \\
\hline \hline
\end{tabular}
\end{table}

\subsection{Channel transmittances} 
The relationship between the communication distance $L_{a(b)}$ and the channel transmittance $\eta^{a(b)}$ is
\begin{equation} \label{eq20}
    \begin{split}
        L_{a(b)}=\frac{10\lg\frac{\eta_{d}}{\eta^{a(b)}}}{\alpha},
    \end{split}
\end{equation}
where the parameters are displayed in Table \ref{tab1}.
Note that $L_{a(b)}$ and $\eta^{a(b)}$ are in one-to-one correspondence.
The difference between the communication distances from Alice and Bob to Charlie ($L_{b}-L_{a}$) depends only on the transmittance ratio $\eta^{a}/\eta^{b}$.
Therefore, when $L_{a}$ and $\eta^{a}/\eta^{b}$ are set as constants, one can directly calculate the value of $L_{b}$.

Variations in both channel transmittances and the maximal pairing interval impact the optimal intensities. 
Therefore, to examine the effect of channel transmittances independently, fixing the maximal pairing interval $\lambda$ is essential.
It is natural to focus on two limiting cases, i.e., $\lambda \to +\infty$ and $\lambda=1$.

When $\lambda \to +\infty$, we derive the following relationships between $\mu^{a}_{m}$ and $\mu^{b}_{m}$ by employing the calculation method described above:
\begin{equation} \label{lam0}
    \begin{split}
        &\mu^{a}_{m}+\mu^{b}_{m}\approx1, \\&\frac{\mu^{b}_{m}}{\mu^{a}_{m}}\approx\sqrt{\delta}=\sqrt{\frac{\eta^{a}}{\eta^{b}}},
    \end{split}
\end{equation}
where these approximations are due to the dark count rate $p_{d}$ and the approximation error of the Taylor series. 
If these errors are not considered, the approximations become the equalities.
The detailed derivation of these relationships is given in Appendix \ref{appB}.
Note that the above relationships for the optimal intensities differ from that of the asymmetric case for other protocols \cite{wang2019asymmetric, grasselli2019asymmetric, wang2020simple, zhou2019asymmetric}.
This is because the intensities of the $Z$ and $X$ bases in asymmetric MP-QKD are coupled.

In the case of $\lambda\to +\infty$, the optimal pulse intensities depend on the transmittance ratio $\eta^{a}/\eta^{b}$.
Therefore, once the difference between the communication distances ($L_{b}-L_{a}$) is determined, one can approximately derive the optimal pulse intensities ($\mu^{a}_{m}$ and $\mu^{b}_{m}$) using the above relationship.

\begin{figure}[htbp]
\centering
\includegraphics[scale=0.46]{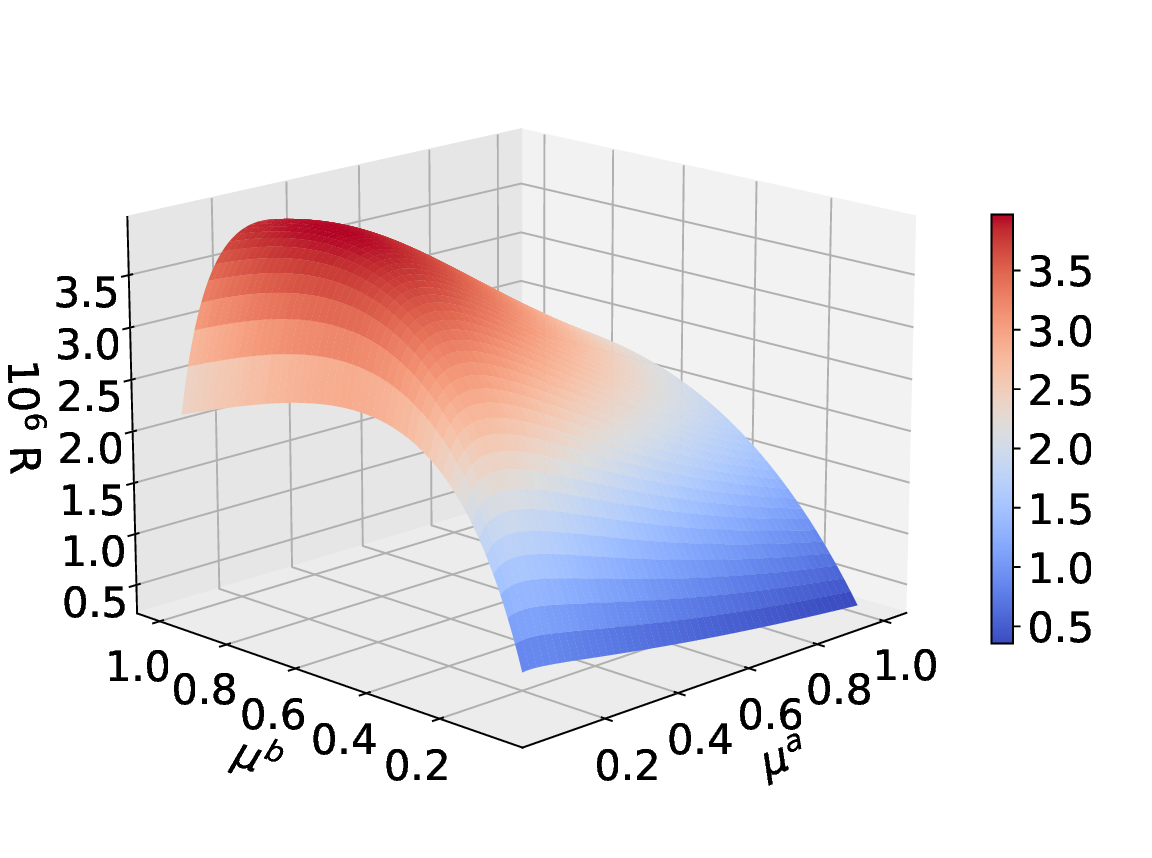}
\caption{An example of key rate $R$ versus two pulse intensities ($\mu^{a}$ and $\mu^{b}$). 
The communication distance from Alice to Charlie is set to $L_{a}=100\ \text{km}$, the transmittance ratio is fixed to $\eta^{a}/\eta^{b}=10$ (equivalent to $L_b = 150\ \text{km}$), and the maximal pairing interval is set to $\lambda=10^{6}$.
The pulse intensities that maximize the key rate are $\mu^{a}_{m}=0.2402$ and $\mu^{b}_{m}=0.7594$, respectively, and their corresponding ratios are $\mu^{b}_{m}/\mu^{a}_{m}=3.1615$.}
\label{fig_single}
\end{figure}

In Table \ref{tab_lam0}, we calculate the optimal pulse intensities for various ratios of channel transmittances at $\lambda=10^{6}$. 
$\lambda=10^{6}$ is selected to approximate $\lambda\to+\infty$.
The errors are not ignored in the calculation process.
Note that $\eta^{a}/\eta^{b}$ and ($L_{b}-L_{a}$) share a one-to-one correspondence.
For the simplicity of discussion, the communication distance from Alice to Charlie
is set to $L_{a}=100\ \text{km}$, and the difference between two communication distances is defined as $\Delta:=L_{b}-L_{a}$.
Clearly, $\mu^{a}_{m}$ and $\mu^{b}_{m}$ closely approximate Eq. (\ref{lam0}) when $\lambda=10^{6}$.
Moreover, we plot a three-dimensional image with a transmittance ratio of $\eta^{a}/\eta^{b}=10$ in Fig. \ref{fig_single}. 
It is observed that only one set of $\mu^{a}_{m}$ and $\mu^{b}_{m}$ allows the key rate to reach its peak value.
This can be straightforwardly derived by analyzing monotonicity and concavity.
Note that variations in $\lambda$ and $\eta^{a}/\eta^{b}$ do not impact the uniqueness of the optimal intensities.

\begin{table}[htbp]
\caption{Example comparison of optimal pulse intensities for different channel-transmittance ratios at $\lambda=10^{6}$. 
The maximal pairing interval is set to $\lambda=10^{6}$, approximating the case where $\lambda\to+\infty$.
The communication distance from Alice to Charlie is set to $L_{a}=100\ \text{km}$.
The difference between the communication distances from Alice and Bob to Charlie is defined as $\Delta:=L_{b}-L_{a}$. 
To ensure precision, we maintain intensity values to four decimal places.}
\label{tab_lam0}
\begin{tabular}{p{1cm}<{\centering} p{1.2cm}<{\centering} p{1.5cm}<{\centering} p{1.5cm}<{\centering} p{1.5cm}<{\centering}}
\hline \hline
\noalign{\vskip 1pt} 
$\Delta$ & $\eta^{a}/\eta^{b}$ & $\mu^{a}_{m}$ & $\mu^{b}_{m}$ & $\mu^{b}_{m}/\mu^{a}_{m}$ \\ 
\hline
0 & 1 & 0.4998 & 0.4998 & 1 \\
50 & 10 & 0.2402 & 0.7594 & 3.1615 \\ 
100 & 100 & 0.0901 & 0.9011 & 10.0011 \\ 
\hline \hline
\end{tabular}
\end{table}

Next, we discuss the case of $\lambda=1$. 
By using the calculation method, we conclude that when the value of $\eta^{a}/\eta^{b}$ is relatively small, the optimal pulse intensities are approximated as
\begin{equation} \label{lam1}
    \begin{split}
        \mu^{a}_{m}\approx1,\quad \mu^{b}_{m}\approx1.
    \end{split}
\end{equation}
The approximations mentioned here differ from those in Eq. (\ref{lam0}).
The approximations in Eq. (\ref{lam1}) are influenced by not just the dark count rate $p_{d}$ and the error term of the Taylor series but also by the probability of a successful detector click $p$.
The detailed derivation of this conclusion is given in Appendix \ref{appB}.
Note that when the value of $\eta^{a}/\eta^{b}$ is large, $\mu^{a}_{m}$ and $\mu^{a}_{m}$ in Eq. (\ref{lam1}) yield bias.

In Table \ref{tab_lam1}, we determine the optimal pulse intensities for different channel-transmittance ratios at $\lambda=1$.
Errors are taken into account throughout the calculation process.
When $\eta^{a}/\eta^{b}$ is relatively small, both $\mu^{a}_{m}$ and $\mu^{a}_{m}$ satisfy Eq. (\ref{lam1}) well. 
However, when $\eta^{a}/\eta^{b}$ is large, both $\mu^{a}_{m}$ and $\mu^{a}_{m}$ deviate from 1, with $\mu^{a}_{m}$ deviating more significantly.

\begin{table}[htbp]
\caption{Example comparison of optimal pulse intensities for different channel-transmittance ratios at $\lambda=1$. 
The communication distance from Alice to Charlie is set to $L_{a}=100\ \text{km}$. 
The difference between the communication distances from Alice and Bob to Charlie is defined as $\Delta:=L_{b}-L_{a}$. 
To ensure precision, we maintain intensity values to four decimal places.}
\label{tab_lam1}
\begin{tabular}{p{1cm}<{\centering} p{1.2cm}<{\centering} p{1.5cm}<{\centering} p{1.5cm}<{\centering}}
\hline \hline
\noalign{\vskip 1pt} 
$\Delta$ & $\eta^{a}/\eta^{b}$ & $\mu^{a}_{m}$ & $\mu^{b}_{m}$ \\ 
\hline
0 & 1 & 0.9962 & 0.9962 \\
50 & 10 & 0.9802 & 0.9962 \\ 
100 & 100 & 0.8682 & 0.9812 \\ 
\hline \hline
\end{tabular}
\end{table}

\subsection{Maximal pairing interval}
The maximal pairing interval $\lambda$ has an impact on the pairing ratio $r_{p}(p,\lambda)$, which is calculated as
\begin{equation} \label{eq22}
    \begin{split}
        r_{p}(p,\lambda)=\left \{\frac{1}{p\left [1-(1-p)^{\lambda}\right]}+\frac{1}{p}\right \}^{-1},
    \end{split}
\end{equation}
where $p$ is the probability of a successful detector click, given approximately by $(\eta^{a}\mu^{a}+\eta^{b}\mu^{b})/2$ in asymmetric MP-QKD.
The derivation of this formula is shown in Ref. \cite{zeng2022mode}.

If the maximal interval is set to $\lambda\to+\infty$, then
\begin{equation} \label{rp1}
    \begin{split}
        r_{p}=\frac{p}{2}\approx \frac{\eta^{a}\mu^{a}+\eta^{b}\mu^{b}}{4}.
    \end{split}
\end{equation}
On the other hand, if $\lambda=1$, then
\begin{equation} \label{rp2}
    \begin{split}
        r_{p}=\frac{p^{2}}{1+p}\approx\frac{(\eta^{a}\mu^{a}+\eta^{b}\mu^{b})^2}{4+2(\eta^{a}\mu^{a}+\eta^{b}\mu^{b})}.
    \end{split}
\end{equation}

When $\lambda$ takes values between 1 and positive infinity, the impact of channel transmittances on the optimal intensities exhibits a trend.

We explore the influence of varying the maximal pairing interval $\lambda$ on the optimal intensities in both the symmetric and asymmetric cases, i.e., when $\eta^{a}/\eta^{b}=1$ and when $\eta^{a}/\eta^{b}\ne 1$.
The dark count rate and the error term of the Taylor series are not ignored in the calculation process.

In the asymmetric case, for the simplicity of discussion, we assume $L_{a}=100\ \text{km}$ and $\eta^{a}/\eta^{b}=10$.
We then calculate the optimal intensities at different maximal intervals, as shown in Table \ref{tab_asy}.
As $\lambda$ increases from 1 to $10^{6}$, the sum of the optimal intensities ($\mu^{a}_{m}+\mu^{b}_{m}$) approximates a progressive decrease from 2 to 1, and their ratio ($\mu^{b}_{m}/\mu^{a}_{m}$) approximates a gradual increase from 1 to $\sqrt{10}$.
This trend corresponds to the transition from Eq. (\ref{lam1}) to Eq. (\ref{lam0}).

\begin{table}[htbp]
\caption{Example comparison of optimal pulse intensities for different maximal pairing intervals in the asymmetric case. 
We set $L_{a}=100\ \text{km}$ and $\eta^{a}/\eta^{b}=10$ (equivalent to $L_b = 150\ \text{km}$).
To ensure precision, we maintain intensity values to four decimal places.}
\label{tab_asy}
\begin{tabular}{p{1cm}<{\centering} p{1.5cm}<{\centering} p{1.5cm}<{\centering} p{1.5cm}<{\centering} p{1.5cm}<{\centering}}
\hline \hline
\noalign{\vskip 1pt} 
$\lambda$ & $\mu^{a}_{m}$ & $\mu^{b}_{m}$ & $\mu^{b}_{m}/\mu^{a}_{m}$ \\ 
\hline
1 & 0.9802 & 0.9962 & 1.0163 \\
$10^{1}$ & 0.9677 & 0.9952 & 1.0284 \\
$10^{2}$ & 0.8707 & 0.9838 & 1.1299 \\
$10^{3}$ & 0.5512 & 0.9239 & 1.6761 \\ 
$10^{4}$ & 0.2687 & 0.7851 & 2.9218 \\
$10^{5}$ & 0.2399 & 0.7592 & 3.1647 \\
$10^{6}$ & 0.2402 & 0.7594 & 3.1615 \\ \hline \hline
\end{tabular}
\end{table}

In the symmetric case ($\eta^{a}/\eta^{b}=1$), we set $L_{a}=100\ \text{km}$ and calculate the optimal intensities at various maximal intervals, as presented in Table \ref{tab_sy}.
As $\lambda$ ranges from $1$ to $10^{6}$, the optimal intensities $\mu^{a}_{m}$ and $\mu^{b}_{m}$ both approximate a decline from 1 to 0.5, corresponding to the transition from Eq. (\ref{lam1}) to Eq. (\ref{lam0}). 
This trend is consistent with observations in the asymmetric case.

\begin{table}[htbp]
\caption{Example comparison of optimal pulse intensities for different maximal pairing intervals in the symmetric case. 
We set $L_{a}=100\ \text{km}$ and $\eta^{a}/\eta^{b}=1$ (equivalent to $L_b = 100\ \text{km}$).
To ensure precision, we maintain intensity values to four decimal places.}
\label{tab_sy}
\begin{tabular}{p{1cm}<{\centering} p{1.5cm}<{\centering} p{1.5cm}<{\centering} p{1.5cm}<{\centering} p{1.5cm}<{\centering}}
\hline \hline
\noalign{\vskip 1pt} 
$\lambda$ & $\mu^{a}_{m}$ & $\mu^{b}_{m}$ & $\mu^{b}_{m}/\mu^{a}_{m}$ \\ 
\hline
1 & 0.9962 & 0.9962 & 1 \\
$10^{1}$ & 0.9838 & 0.9838 & 1 \\
$10^{2}$ & 0.8915 & 0.8915 & 1 \\
$10^{3}$ & 0.6424 & 0.6424 & 1 \\ 
$10^{4}$ & 0.5008 & 0.5008 & 1 \\
$10^{5}$ & 0.5005 & 0.5005 & 1 \\
$10^{6}$ & 0.4998 & 0.4998 & 1 \\ 
\hline \hline
\end{tabular}
\end{table}

The reason for this trend is that the final key rate $R$ is influenced by parameters such as the pairing ratio, the probability of successful detection, and the error rate.
Variation in $\lambda$ results in changes to the weight of the pairing ratio $r_{p}(p,\lambda)$ in $R$ [e.g., Eqs. (\ref{rp1}) and (\ref{rp2})], consequently impacting the optimal intensities ($\mu^{a}_{m}$ and $\mu^{b}_{m}$).

\section{Numerical Simulations} \label{sec5}
In this section, we plot the optimal pulse intensities for two representative cases and simulate the asymptotic performance of asymmetric MP-QKD in different scenarios.
Note that the dark count rate and the approximation error of the Taylor series are considered throughout the simulations.

\begin{figure}[htbp]
\centering
\includegraphics[scale=0.305]{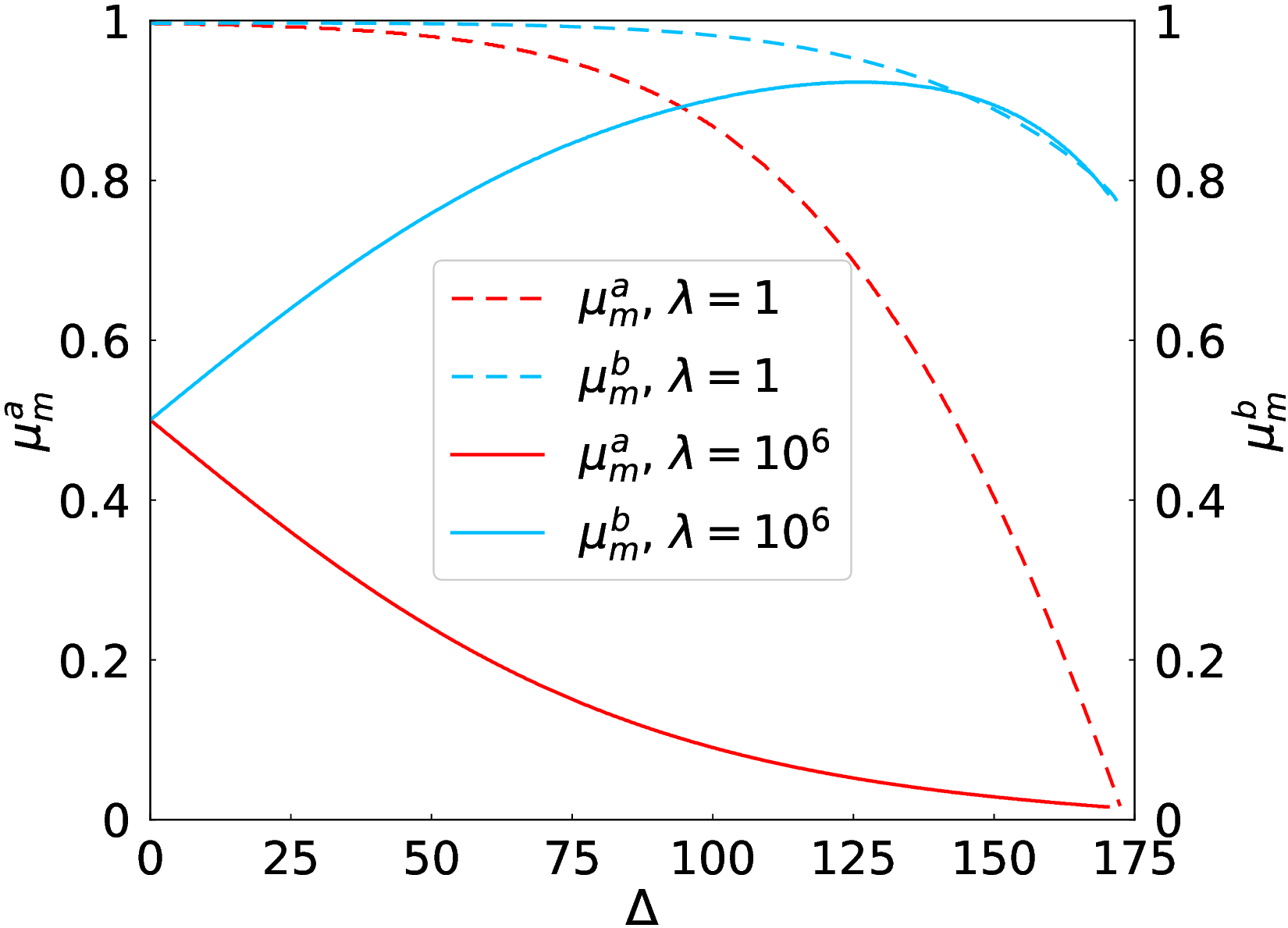}
\caption{Optimal pulse intensities ($\mu^{a}_{m}$ and $\mu^{b}_{m}$) versus the difference between the two distances $\Delta$ at $\lambda=10^6$ and $\lambda=1$.
$\lambda=10^{6}$ is selected to approximate $\lambda\to+\infty$.
For simplicity, the communication distance from Alice to Charlie
is set to $L_{a}=100\ \text{km}$, and the difference between two communication distances is defined as $\Delta:=L_{b}-L_{a}$.
Solid lines indicate $\lambda=10^{6}$, while dashed lines represent $\lambda=1$.
In both line types, $\mu^{a}_{m}$ corresponds to a lower line position than $\mu^{b}_{m}$.
}
\label{fig_opt}
\end{figure}

In Fig. \ref{fig_opt}, the optimal pulse intensities are plotted as a function of the difference between the two distances for $\lambda=10^{6}$ and $\lambda=1$. Here, $\lambda=10^{6}$ is selected to approximate $\lambda\to+\infty$.
For the simplicity of discussion, the communication distance from Alice to Charlie is set to $L_{a}=100\ \text{km}$.
As $\Delta$ incrementally increases, $\mu^{a}_{m}$ and $\mu^{b}_{m}$ at $\lambda=10^6$ conform to Eq. (\ref{lam0}). 
Conversely, $\mu^{a}_{m}$ and $\mu^{b}_{m}$ at $\lambda=1$ gradually deviate from an initial approximation of 1, where the deviation of $\mu^{a}_{m}$ is more pronounced.
Furthermore, when $\Delta$ is significantly large, the relationships between $\mu^{a}_{m}$ and $\mu^{b}_{m}$ at 
$\lambda=10^6$ similarly deviate.
These deviations emerge because, with the increase in $\Delta$, the impact of the dark count rate $p_{d}$ becomes progressively pronounced.
Note that when $\Delta$ is large enough, $\lambda$ has a negligible effect on $\mu^{a}_{m}$ and $\mu^{b}_{m}$.

When dealing with the practical challenge of asymmetric MP-QKD, apart from utilizing the optimal-pulse-intensity method, one can also equalize the distance discrepancy by adding extra fiber on the closer side so that the transmittances of both sides become uniform.
However, a drawback of this approach is that it reduces the total transmittance, ultimately leading to a smaller key rate.

\begin{figure}[htbp]
\centering
\includegraphics[scale=0.345]{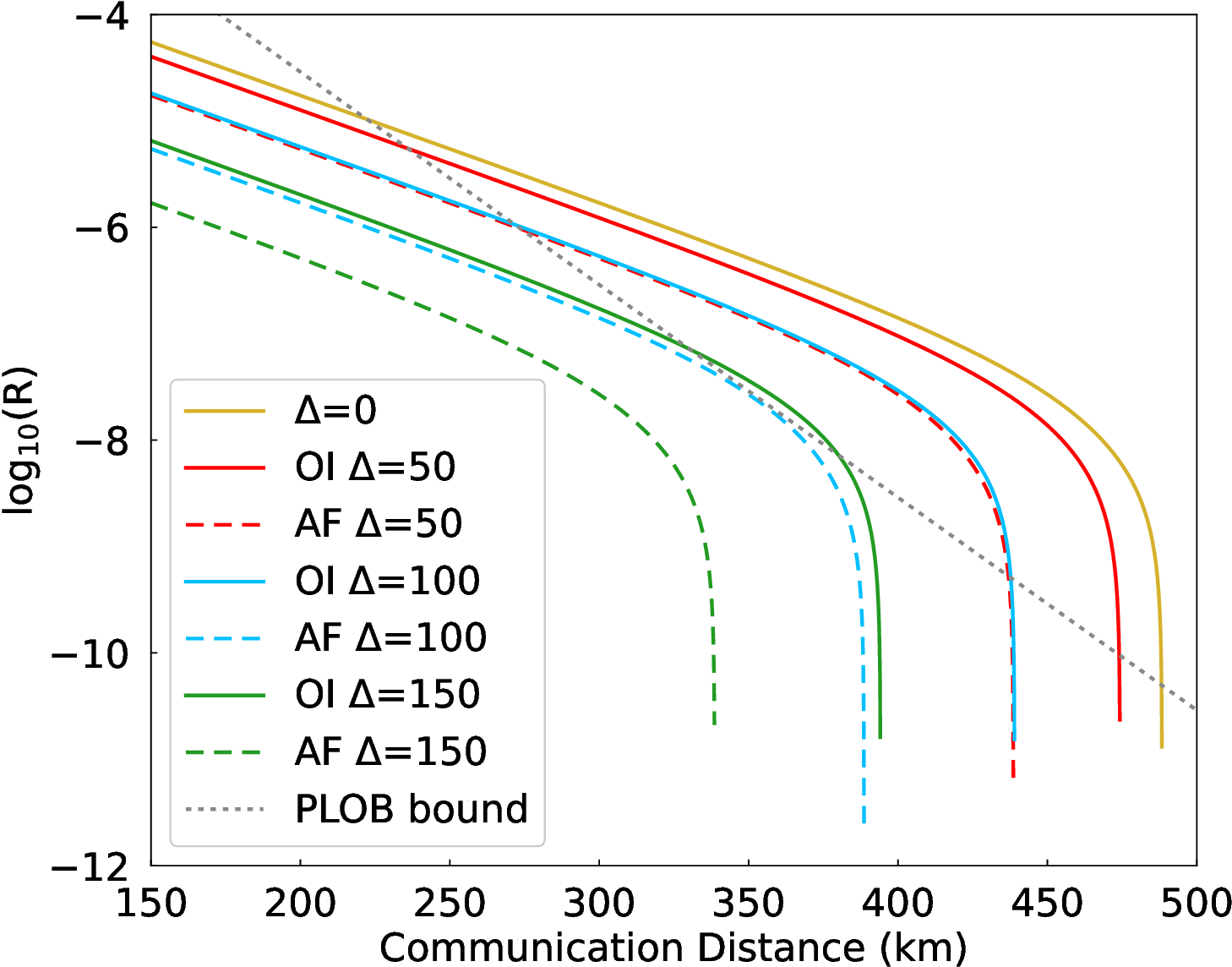}
\caption{Simulation plot of the final key rate $R$ versus the total communication distance ($L_{a}+L_{b}$) at $\lambda=10^{6}$ for different methods.  
The difference between the communication distances from Alice and Bob to Charlie is defined as $\Delta:=L_{b}-L_{a}$. 
$\Delta=0$ signifies the original symmetric MP-QKD, represented by the solid (top) line.
“OI” refers to the optimal intensity method, depicted by the solid line.
“AF” denotes the adding-fiber method, shown as the dashed line.
“PLOB bound” represents the repeaterless rate-transmittance bound, illustrated by the dotted line.
In the OI and AF scenarios, a larger value of $\Delta$ corresponds to a lower line position.}
\label{fig_dif_0}
\end{figure}

The simulation result for MP-QKD using different methods is shown in Fig. \ref{fig_dif_0}. 
The maximal interval is set to $\lambda=10^{6}$.
It can be observed that by employing the optimal intensity method, one can achieve higher key rates compared to the adding-fiber method for the same difference $\Delta$.
As the difference $\Delta$ increases, the final key rate $R$ decreases.
Nevertheless, even when $\Delta=150\ \text{km}$, the key rate using optimal intensities can surpass the PLOB bound at a total communication distance of around 350 km.

\begin{figure}[htbp]
\centering
\includegraphics[scale=0.325]{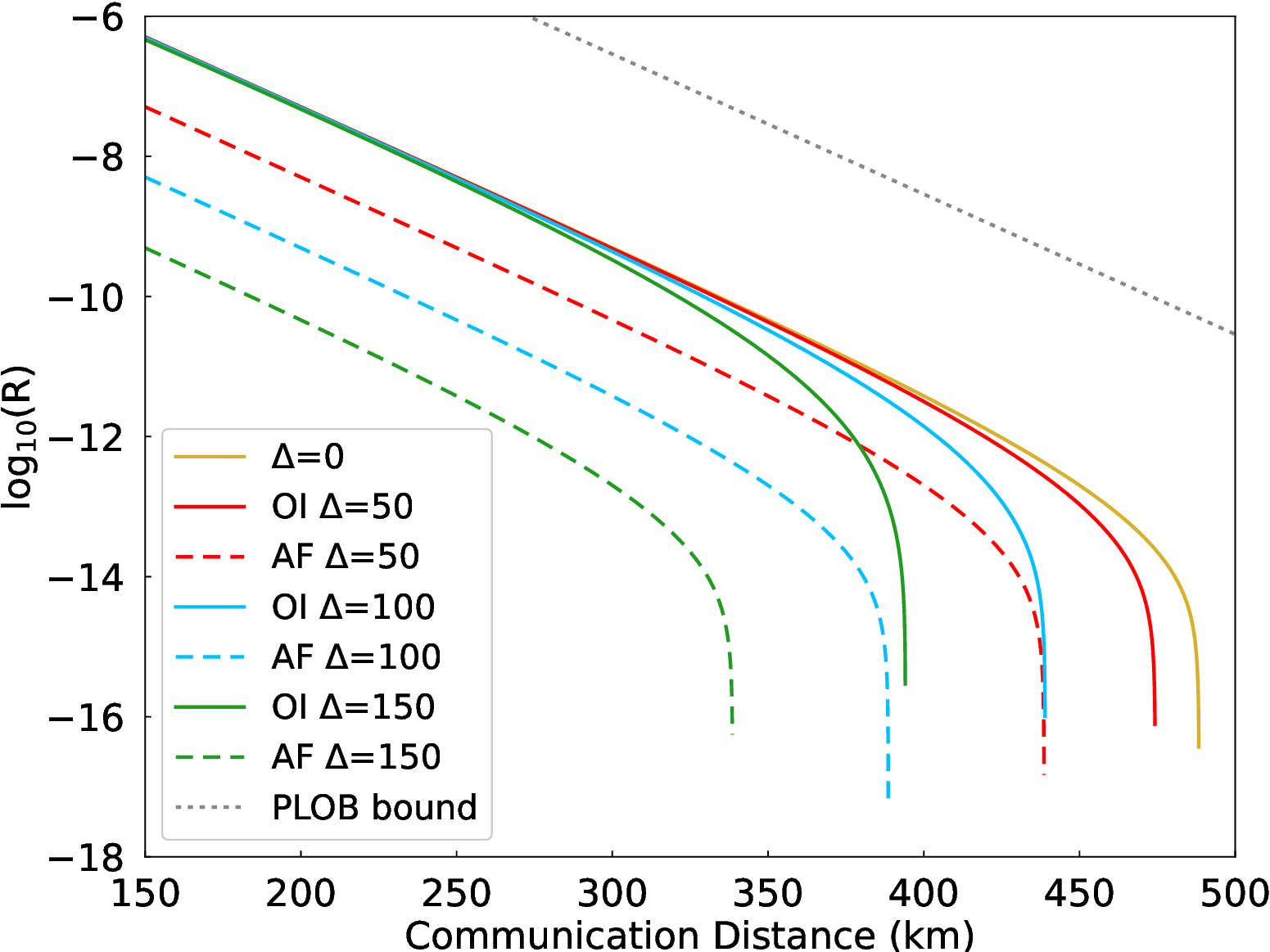}
\caption{Simulation plot of the final key rate $R$ versus the total communication distance ($L_{a}+L_{b}$) at $\lambda=1$ for different methods. 
$\Delta=0$ signifies the original symmetric MP-QKD, represented by the solid (top) line.
“OI” refers to the optimal intensity method, depicted by the solid line.
“AF” denotes the adding fiber method, shown as the dashed line.
“PLOB bound” represents the repeaterless rate-transmittance bound, illustrated by the dotted line.
In the OI and AF scenarios, a larger value of $\Delta$ corresponds to a lower line position.}
\label{fig_dif_1}
\end{figure}

Moreover, in Fig. \ref{fig_dif_1}, the performance of MP-QKD at $\lambda=1$ using different methods is simulated.
Similarly, utilizing the optimal intensity method results in higher key rates than the adding-fiber method for the same $\Delta$.
For the same method, an increase in $\Delta$ corresponds to a decrease in $R$.

\begin{figure}[htbp]
\centering
\includegraphics[scale=0.35]{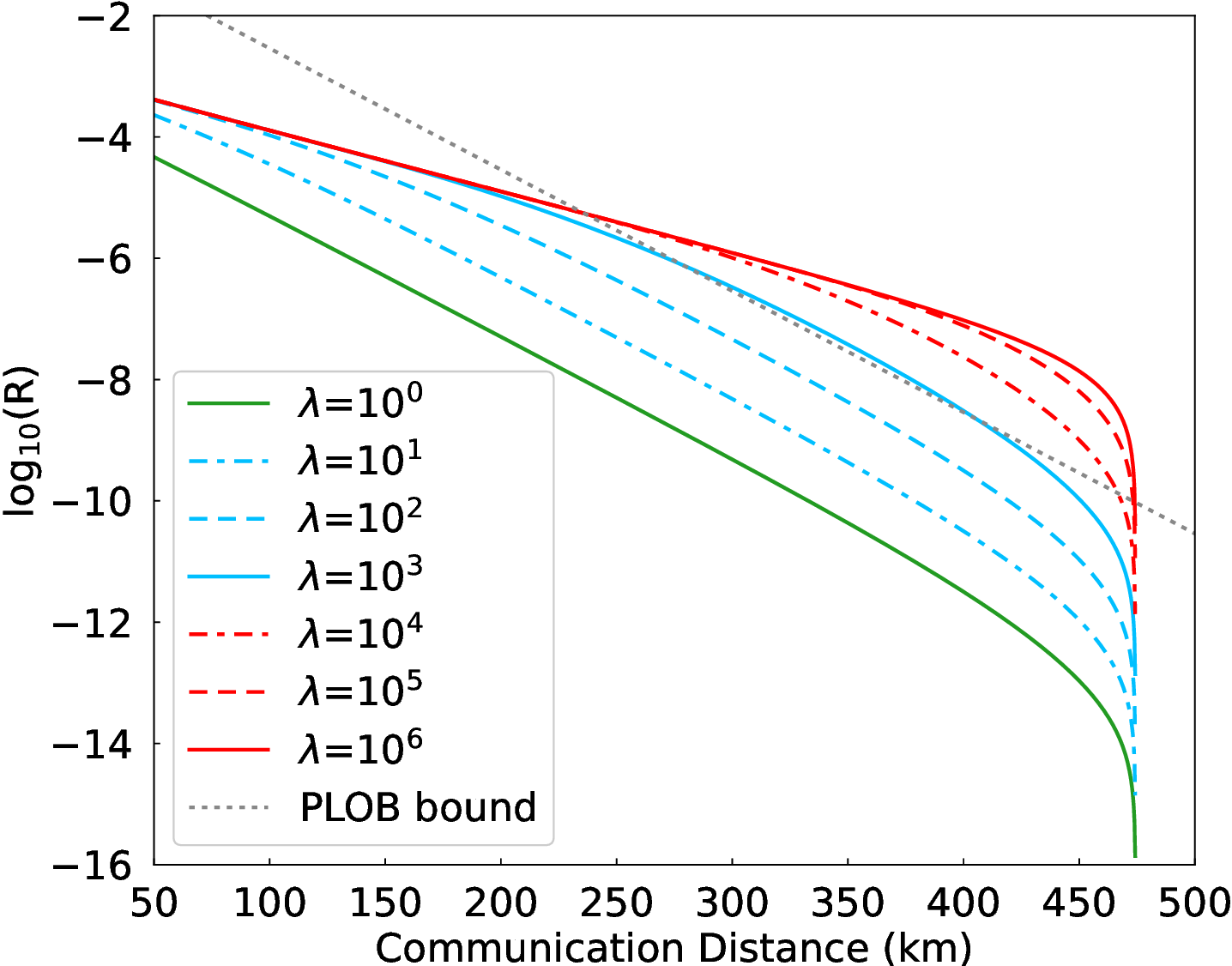}
\caption{Simulation of the final key rate $R$ versus communication distance ($L_{a}+L_{b}$) at different maximal pairing intervals $\lambda$.
The difference between two communication distances is set to $\Delta=50\ \text{km}$.
The optimal-pulse-intensity method is used for each case.
A smaller value of $\lambda$ corresponds to a lower line position.}
\label{fig_interval}
\end{figure}

Figure \ref{fig_interval} illustrates the relationship between rate and distance at different maximal pairing intervals employing the optimal intensity method, where $\Delta=50\ \text{km}$.
It can be observed that the variation in $\lambda$ does not influence the maximal communication distance.
The key rate $R$ gradually increases as maximal interval $\lambda$ rises, and it approaches saturation when $\lambda=10^{6}$.
When $\lambda$ is increased to 1000, the key rate experiences a substantial enhancement, surpassing the $\lambda=1$ case by three orders of magnitude, and it exceeds the repeaterless rate-transmittance bound at a total communication distance of approximately 300 km.

\begin{figure}[htbp]
\centering
\includegraphics[scale=0.38]{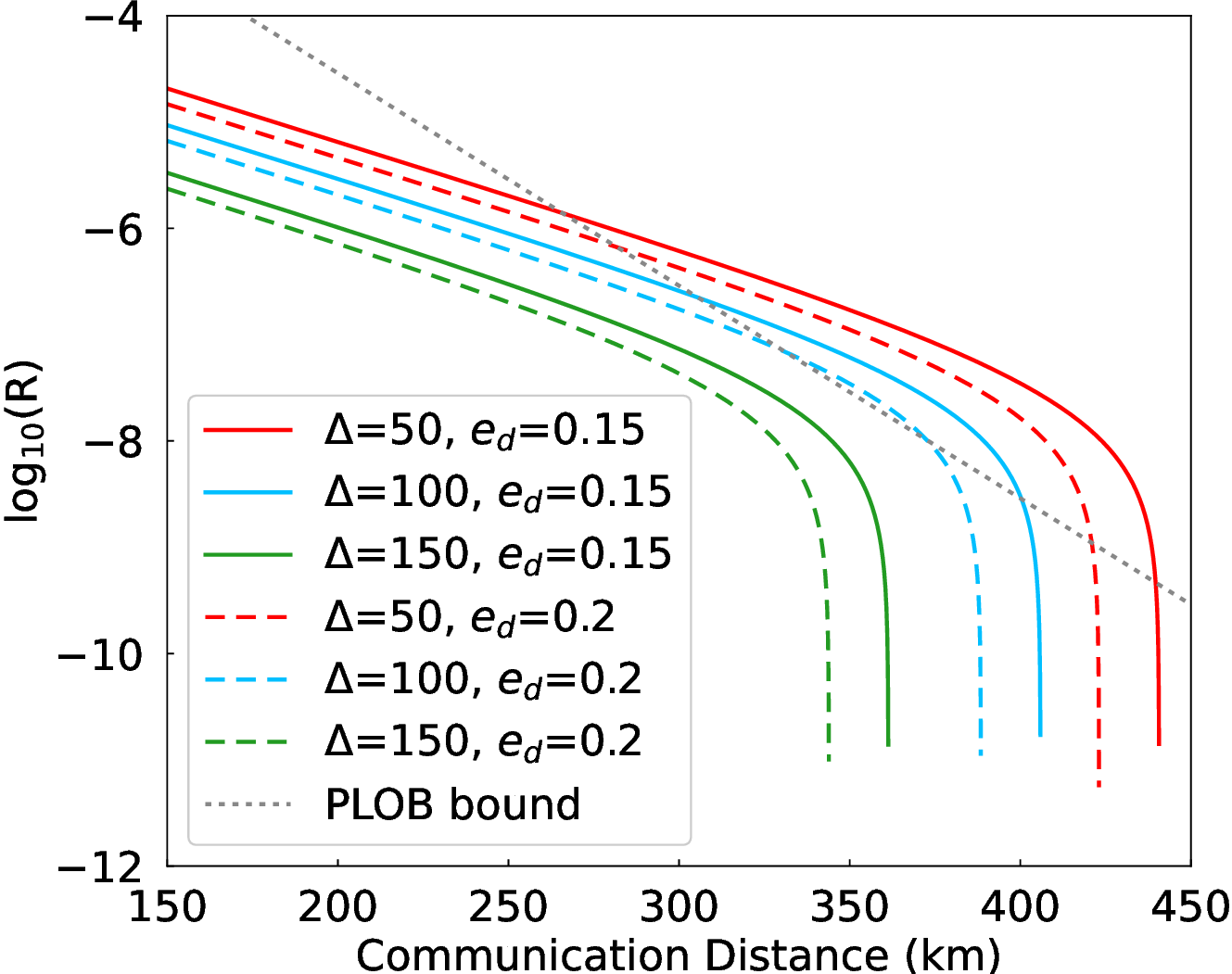}
\caption{Final key rate $R$ as a function of communication distance $(L_{a}+L_{b})$ at different misalignment error rates $e_{d}$. 
The difference between the communication distances from Alice and Bob to Charlie is defined as $\Delta:=L_{b}-L_{a}$.
The maximal pairing interval is set to $\lambda=10^{6}$.
The optimal-pulse-intensity method is used for each case.
Within the same line type, a larger value of $\Delta$ corresponds to a lower line position.
}
\label{fig_error}
\end{figure}

To demonstrate the tolerance of misalignment errors in asymmetric MP-QKD using optimal pulse intensities, we present simulation results depicting the key rate $R$ versus communication distance at varying misalignment error rates $e_{d}$ in Fig. \ref{fig_error}.
For simplicity, the maximal pairing interval is set to $\lambda=10^{6}$.
The results show that when employing the optimal-pulse-intensity method, asymmetric MP-QKD exhibits remarkable robustness against misalignment errors. 
Even with the difference of $\Delta=100$, this scheme can still surpass the PLOB bound when the misalignment error rate is $e_{d}=20\%$.

\section{Conclusion} \label{sec6}
In this paper, we extended the application of MP-QKD to the asymmetric case, which substantially enhances the protocol's utility.

First, we outlined the steps involved in asymmetric MP-QKD. 
We then analyzed the security of the protocol using decoy-state estimation.
It is important to note that asymmetric intensities and asymmetric channel transmittances do not impact the security of this protocol.

Second, the performance of the protocol in the asymmetric case can be improved by selecting the appropriate pulse intensities.
However, the intensity relationships among modes in the asymmetric MP-QKD differ from those in previous asymmetric protocols because the intensities between different modes in MP-QKD cannot be decoupled. 
For this, we introduced an innovative optimal-pulse-intensity method, which can enhance key rates by determining ideal pulse intensities.
We summarized the relationships and trends of optimal intensities at different channel transmittances and maximal pairing intervals.
Therefore, we illustrated how the appropriate pulse intensities can be chosen to optimize the key rate in various asymmetric MP-QKD scenarios.

Furthermore, we plotted the optimal pulse intensities for two representative cases.
We conducted a comparative assessment of the optimal-pulse-intensity method against the approach of adding additional fibers.
We simulated the performance of asymmetric MP-QKD under different conditions using the optimal intensity method. 
Additionally, we displayed the tolerance of misalignment errors in asymmetric MP-QKD.
Our simulation results clearly indicate that the optimal-pulse-intensity method not only is practical but also can significantly mitigate the effects of asymmetric communication distances on protocol performance.

Finally, in future research, we will focus on the statistical analysis of the finite key scenario for asymmetric MP-QKD. 
This analysis could build on insights from a recent study that investigated the implications of finite key length in the original MP-QKD, which was enhanced using a six-state method \cite{wang2023tight}.
Moreover, it would be interesting to combine asymmetric MP-QKD with many advanced quantum technologies, such as advantage distillation \cite{tan2020advantage}, photonic graph states \cite{huang2023chip}, quantum squeezing \cite{wang2023squeezing, lu2023recent}, and optical clocks \cite{zhang2023development}. 
That would further improve the performance and practicality of the protocol.

\begin{acknowledgments}
This work was supported by the National Natural Science Foundation of China (Basic Science Center Program: No. 61988101), the National Natural Science Foundation of China under Grant No. 12105105, the Natural Science Foundation of Shanghai under Grant No. 21ZR1415800, the Shanghai Sailing Program under Grant No. 21YF1409800, the Programme of Introducing Talents of Discipline to Universities (the 111 Project) under Grant No. B17017, and the startup fund from East China University of Science and Technology under Grant No. YH0142214.
\end{acknowledgments}

\appendix
\section{Simulation details} \label{appA}
In this appendix, we derive the simulation expression for the asymmetric MP-QKD.
The final key rate of the asymmetric MP-QKD is given in Eq. (\ref{eq18}).
We elaborate on each parameter in this formula.

In the asymptotic case, it is assumed that Alice (Bob) randomly prepares pulses with intensity $\{ 0,\mu^{a}\}$ ($\{ 0,\mu^{b}\}$), each having a probability of approximately $1/2$, while assigning a negligible probability to the decoy intensity $\nu^{a}$ ($\nu^{b}$).
For the simplicity of discussion, we denote the coherent pulse emitted by Alice (Bob) in the $i$ th round as $|\sqrt{z^{a}_{i}\mu^{a}} e^{i\phi^{a}_{i}}\rangle$ ($|\sqrt{z^{b}_{i}\mu^{b}} e^{i\phi^{b}_{i}}\rangle$), where $z^{a(b)}_{i}\in\{0,1\}$ is a random variable signifying the intensity and $\phi^{a(b)}_{i}\in[0,2\pi)$ is a random phase.
We denote the intensity setting for the $i$ th round using the vector 
\begin{equation}
    \begin{split}
        z_{i}:=[z^{a}_{i},z^{b}_{i}]\in\{00,01,10,11\}.
    \end{split}
\end{equation}

In the practical asymmetric MP-QKD, Alice and Bob transmit pulses to Charlie through two asymmetric loss channels. 
These two channel transmittances are $\eta^{a}$ and $\eta^{b}$, respectively.
The relevant simulation data are listed in Table \ref{tab1}.
The channel is independent and identically distributed (i.i.d.) for each round.
Alice and Bob proceed by pairing the successful pulses and sifting their bases.
To pair the $(i,j)$ th pulses, we define $\tau_{i,j}=[\tau^{a}_{i,j},\tau^{b}_{i,j}]:=[z^{a}_{i}\oplus z^{a}_{j}, z^{b}_{i}\oplus z^{b}_{j}]$, where $\oplus$ represents the bitwise addition modulo 2.
When $\tau_{i,j}=[1,1]$, the $(i,j)$ pair is set to be an effective $Z$ pair.

In the $i$ th round, the click events of the left and right detectors can be denoted as two variables $(L_{i},R_{i})$.
The detector click variable is defined as $C_{i}:=L_{i}\oplus R_{i}$.
When $C_{i}=1$, it signifies the occurrence of a successful click.
The detection probability $\text{Pr}(C_{i}=1|z_{i})$ is
\begin{equation} \label{A2}
    \begin{split}
        \text{Pr}(C_{i}=1|z_{i})\approx 1-(1-2p_{d})e^{-\eta^{a}\mu^{a}z^{a}_{i}-\eta^{b}\mu^{b}z^{b}_{i} }.
    \end{split}
\end{equation}
The successful click probability of each round is
\begin{equation}
    \begin{split}
        p&:=\text{Pr}(C_{i}=1)=\sum_{z_{i}}\text{Pr}(C_{i}=1|z_{i})\text{Pr}(z_{i})\\&=\frac{1}{4}\sum_{z_{i}}\text{Pr}(C_{i}=1|z_{i}).
    \end{split}
\end{equation}
One can then calculate the pairing rate $r_{p}(p,\lambda)$ via Eq. (\ref{eq22}).

The coherent states emitted in the $i$ th round can be viewed as a linear superposition of photon-number states.
Suppose the photon number for the $i$ th round is $n_{i}:=[n^{a}_{i},n^{b}_{i}]$.
Given photon-number state $|n_{i}\rangle$ emitted by Alice and Bob, the detection probability $\text{Pr}(C_{i}=1|n_{i})$ is expressed as
\begin{equation}
    \begin{split}
        \text{Pr}(C_{i}=1|n_{i})\approx 1-(1-2p_{d})(1-\eta^{a})^{n^{a}_{i}}(1-\eta^{b})^{n^{b}_{i}}.
    \end{split}
\end{equation}

Without loss of generality, we regard the $i$ th and $j$ th rounds as a paired group.
Since the detection are i.i.d. for all rounds, the probability for the intensity setting $z_{i(j)}$ caused by a successful click is
\begin{equation}
    \begin{split}
        \text{Pr}(z_{i(j)}|C_{i(j)}=1)&=\frac{\text{Pr}(z_{i(j)},C_{i(j)}=1)}{\text{Pr}(C_{i(j)}=1)}\\&=\frac{\text{Pr}(C_{i(j)}=1|z_{i(j)})\text{Pr}(z_{i(j)})}{\sum_{z'_{i(j)}}\text{Pr}(C_{i(j)}=1|z'_{i(j)})\text{Pr}(z'_{i(j)})}\\&=\frac{\text{Pr}(C_{i(j)}=1|z_{i(j)})}{\sum_{z'_{i(j)}}\text{Pr}(C_{i(j)}=1|z'_{i(j)})}.
    \end{split}
\end{equation}

When $\tau_{i,j}=[1,1]$, a paired group indexed as $i$ and $j$ is viewed as an effective $Z$ pair.
Therefore, four possible combinations of $z_{i}$ and $z_{j}$ are given by
\begin{equation}
    \begin{split}
        [z_{i},z_{j}]\in \left \{[00,11],[01,10],[10,01],[11,00]\right \},
    \end{split}
\end{equation}
where two combinations causing bit errors are defined as $\text{Err}:=\{[00,11],[11,00]\}$.
For the simplicity of discussion, we adopt several concise representations:
\begin{equation}
    \begin{split}
        &\text{Pr}(C)=\text{Pr}(\text{Pair Clicked}):=\text{Pr}(C_{i}=C_{j}=1)=p^{2},\\&\text{Pr}(E)=\text{Pr}(\text{Pair Effective}):=\text{Pr}(z_{i}\oplus z_{j}=11),\\&\text{Pr(Err)}=\text{Pr}(\text{Pair Erroneous}):=\text{Pr}([z_{i},z_{j}]\in \text{Err}),\\&\text{Pr}(S)=\text{Pr}(\text{Single-Photon Pair}):=\text{Pr}(n_{i}\oplus n_{j}=11).
    \end{split}
\end{equation}

In this way, the $Z$-pair ratio $r_{s}$ can be calculated by
\begin{equation}
    \begin{split}
        r_{s}&=\text{Pr}(E|C)=\text{Pr}(z_{i}\oplus z_{j}=11|C_{i}=1,C_{j}=1)\\&=\sum_{z_{i}\oplus z_{j}=11}\text{Pr}(z_{i}|C_{i}=1)\text{Pr}(z_{j}|C_{j}=1)\\&=\sum_{z_{i}\oplus z_{j}=11}\frac{\text{Pr}(C_{i}=1|z_{i})\text{Pr}(z_{i})}{\text{Pr}(C_{i}=1)}\frac{\text{Pr}(C_{j}=1|z_{j})\text{Pr}(z_{j})}{\text{Pr}(C_{j}=1)}\\&=\frac{1}{16}\frac{1}{p^{2}}\sum_{z_{i}\oplus z_{j}=11}\text{Pr}(C_{i}=1|z_{i})\text{Pr}(C_{j}=1|z_{j}).
    \end{split}
\end{equation}

The expected bit error rate of the $Z$ pair $e^{(\mu^{a},\mu^{b}),Z}$ is
\begin{equation}
    \begin{split}
        &e^{(\mu^{a},\mu^{b}),Z}=\text{Pr}(\text{Err}|E,C)\\&=\frac{\text{Pr}(\text{Err},E|C)}{\text{Pr}(E|C)}=\frac{\text{Pr}(\text{Err}|C)}{\text{Pr}(E|C)}\\&=\frac{1}{r_{s}}\text{Pr}([z_{i},z_{j}]\in \text{Err}|C_{i}=C_{j}=1)\\&=\frac{1}{r_{s}}\sum_{[z_{i},z_{j}]\in \text{Err}}\text{Pr}(z_{i}|C_{i}=1)\text{Pr}(z_{j}|C_{j}=1)\\&=\frac{1}{r_{s}}\sum_{[z_{i},z_{j}]\in \text{Err}}\frac{\text{Pr}(C_{i}=1|z_{i})\text{Pr}(z_{i})}{\text{Pr}(C_{i}=1)}\frac{\text{Pr}(C_{j}=1|z_{j})\text{Pr}(z_{j})}{\text{Pr}(C_{j}=1)}\\&=\frac{1}{16}\frac{1}{r_{s}p^{2}}\sum_{[z_{i},z_{j}]\in \text{Err}}\text{Pr}(C_{i}=1|z_{i})\text{Pr}(C_{j}=1|z_{j}).
    \end{split}
\end{equation}
The reason why the third equation holds is that the effective pairing case contains the erroneous pairing case.

The single-photon pair ratio for the effective $Z$ pairs $\bar{q}_{(1,1)}$ is
\begin{equation}
    \begin{split}
        &\bar{q}_{(1,1)}=\text{Pr}(S|E,C)=\frac{\text{Pr}(S,E,C)}{\text{Pr}(E,C)}\\&=\frac{1}{r_{s}p^{2}}\sum_{z_{i},z_{j}}\text{Pr}(S,E,C|z_{i},z_{j})\text{Pr}(z_{i},z_{j})\\&=\frac{1}{16}\frac{1}{r_{s}p^{2}}\sum_{z_{i}\oplus z_{j}=11}\text{Pr}(S,C|z_{i},z_{j})\\&=\frac{1}{16}\frac{1}{r_{s}p^{2}}\sum_{z_{i}\oplus z_{j}=11}\text{Pr}(C|S,z_{i},z_{j})\text{Pr}(S|z_{i},z_{j})\\&=\frac{1}{16}\frac{P_{\mu^{a}}(1)P_{\mu^{b}}(1)}{r_{s}p^{2}}[\sum_{z_{i}\oplus z_{j}=11}\text{Pr}(C_{i}=1|n_{i}=z_{i})\\&\quad\ \times\text{Pr}(C_{j}=1|n_{j}=z_{j})],
    \end{split}
\end{equation}
where $P_{\mu^{a(b)}}(k)$ is the Poisson distribution [e.g., $P_{\mu^{a}}(k)=e^{-\mu^{a}}\frac{(\mu^{a})^{k}}{k!}$].

When the decoy-state estimation achieves the desired result, one can directly estimate the gain and error rate of the $X$ basis using the following equation \cite{ma2012alternative}:
\begin{equation}
    \begin{split}
        Y_{(1,1)}&=(1-p_{d})^{2} [ \frac{\eta^{a}\eta^{b}}{2}+(2\eta^{a}+2\eta^{b}-3\eta^{a}\eta^{b})p_{d}\\&\quad\ +4(1-\eta^{a})(1-\eta^{b})p^{2}_{d} ],\\e_{(1,1)}&=\frac{e_{0}Y_{(1,1)}-(e_{0}-e_{d})(1-p^{2}_{d})\frac{\eta^{a}\eta^{b}}{2}}{Y_{(1,1)}},
    \end{split}
\end{equation}
where the error caused by vacuum pulses is $e_{0}=1/2$ and the misalignment error is set to $e_{d}=4\%$.

\section{Derivation details}\label{appB}
In this appendix, we derive the content of Eqs. (\ref{lam0}) and (\ref{lam1}), which describe the relationships satisfied by the optimal pulse intensities ($\mu^{a}$ and $\mu^{b}$) when $\lambda \to +\infty$ and $\lambda=1$.

The approximations in Eqs. (\ref{lam0}) and (\ref{lam1}), as described in the main text, arise from the dark count rate $p_{d}$ and the approximation error of the Taylor series. 
Note that the approximation error of the Taylor series produces little effect on the optimal pulse intensities.
Additionally, when $\Delta$ is relatively small, the interference resulting from $p_{d}$ is negligible.
For simplicity, we consider these errors to be zero in the following derivation.

When $p_{d}=0$, the detection probability in Eq. (\ref{A2}) is
\begin{equation}
    \begin{split}
        \text{Pr}(C_{i}=1|z_{i})&\approx 1-e^{-\eta^{a}\mu^{a}z^{a}_{i}-\eta^{b}\mu^{b}z^{b}_{i}}\\&\approx \eta^{a}\mu^{a}z^{a}_{i}+\eta^{b}\mu^{b}z^{b}_{i},
    \end{split}
\end{equation}
where the second approximation is due to the error term of the Taylor series.
Since this error term is not considered in the following discussion, $\text{Pr}(C_{i}=1|z_{i})=\eta^{a}\mu^{a}z^{a}_{i}+\eta^{b}\mu^{b}z^{b}_{i}$.
On this basis, the successful click probability of each round is
\begin{equation}
    \begin{split}
        p&=\frac{1}{4}\sum_{z_{i}}\text{Pr}(C_{i}=1|z_{i})\\&=\frac{\eta^{a}\mu^{a}+\eta^{b}\mu^{b}}{2}.
    \end{split}
\end{equation}

Similarly, when the dark count rate and the error term in the Taylor series are not taken into account, the detection probability $\text{Pr}(C_{i}=1|n_{i})$ is
\begin{equation}
    \begin{split}
        \text{Pr}(C_{i}=1|n_{i})= 1-(1-\eta^{a})^{n^{a}_{i}}(1-\eta^{b})^{n^{b}_{i}}.
    \end{split}
\end{equation}

Then, the $Z$-pair ratio $r_{s}$ is 
\begin{equation}
    \begin{split}
        r_{s}&=\frac{1}{16}\frac{1}{p^{2}}\sum_{z_{i}\oplus z_{j}=11}\text{Pr}(C_{i}=1|z_{i})\text{Pr}(C_{j}=1|z_{j})\\&=\frac{\eta^{a}\eta^{b}\mu^{a}\mu^{b}}{8p^{2}}.
    \end{split}
\end{equation}

The expected bit error rate of the $Z$ pair $e^{(\mu^{a},\mu^{b}),Z}$ is
\begin{equation}
    \begin{split}
        e^{(\mu^{a},\mu^{b}),Z}&=\frac{1}{16}\frac{1}{r_{s}p^{2}}\sum_{[z_{i},z_{j}]\in \text{Err}}\text{Pr}(C_{i}=1|z_{i})\text{Pr}(C_{j}=1|z_{j})\\&=0.
    \end{split}
\end{equation}

The single-photon pair ratio for the effective $Z$ pairs $\bar{q}_{(1,1)}$ is
\begin{equation}
    \begin{split}
        \bar{q}_{(1,1)}&=\frac{1}{16}\frac{P_{\mu^{a}}(1)P_{\mu^{b}}(1)}{r_{s}p^{2}}[\sum_{z_{i}\oplus z_{j}=11}\text{Pr}(C_{i}=1|n_{i}=z_{i})\\&\quad\ \times\text{Pr}(C_{j}=1|n_{j}=z_{j})]\\&=\frac{\eta^{a}\eta^{b}\mu^{a}\mu^{b}e^{-\mu^{a}}e^{-\mu^{b}}}{8r_{s}p^{2}}.
    \end{split}
\end{equation}

Moreover, the gain of the $X$ basis $Y_{(1,1)}$ is
\begin{equation}
    \begin{split}
        Y_{(1,1)}&=(1-p_{d})^{2} [ \frac{\eta^{a}\eta^{b}}{2}+(2\eta^{a}+2\eta^{b}-3\eta^{a}\eta^{b})p_{d}\\&\quad\ +4(1-\eta^{a})(1-\eta^{b})p^{2}_{d} ]\\&=\frac{\eta^{a}\eta^{b}}{2}.
    \end{split}
\end{equation}
Hence, the corresponding error rate $e_{(1,1)}$ is
\begin{equation}
    \begin{split}
        e_{(1,1)}&=\frac{e_{0}Y_{(1,1)}-(e_{0}-e_{d})(1-p^{2}_{d})\frac{\eta^{a}\eta^{b}}{2}}{Y_{(1,1)}}\\&=e_{d},
    \end{split}
\end{equation}
where the misalignment error is $e_{d}=4\%$.

\subsection{Proof of Equation (\ref{lam0})} \label{appB1}
In Eq. (\ref{lam0}), the maximal pairing interval is set to $\lambda\to +\infty$.
The corresponding pairing ratio $r_{p}(p,\lambda)$ is 
\begin{equation}
    \begin{split}
        r_{p}(p,\lambda)&=\textstyle\lim_{\lambda\to +\infty}\left \{\frac{1}{p\left [1-(1-p)^{\lambda}\right]}+\frac{1}{p}\right \}^{-1}\\&=\frac{p}{2}.
    \end{split}
\end{equation}

Therefore, the key-rate formula for this case is
\begin{equation}
    \begin{split}
        R&=r_{p}(p,\lambda) r_{s} \left \{ \bar{q}_{(1,1)} \left [ 1-H(e_{(1,1)})\right ]-fH(e^{(\mu^{a},\mu^{b})Z}) \right \}\\&=\frac{1-H(4\%)}{8}\frac{\eta^{a}\eta^{b}\mu^{a}\mu^{b}e^{-\mu^{a}}e^{-\mu^{b}}}{\eta^{a}\mu^{a}+\eta^{b}\mu^{b}}\\&=\frac{\left [1-H(4\%)\right ]\eta^{a}}{8}\frac{\mu^{a}\mu^{b}e^{-\mu^{a}}e^{-\mu^{b}}}{\delta\mu^{a}+\mu^{b}}.
    \end{split}
\end{equation}
The third equation stems from the calculation method outlined in Sec. \ref{subsec1}, where we define the channel-transmittance ratio as $\eta^{a}/\eta^{b}=\delta\geq 1$, which is a constant.
Moreover, since $L_{a}$ is fixed as a constant and directly corresponds to $\eta^{a}$ on a one-to-one basis, $\eta^{a}$ is likewise constant.

Next, we aim to determine $\mu^{a}_{m}$ and $\mu^{b}_{m}$, which represent the values of $\mu^{a}$ and $\mu^{b}$ when the function $R$ reaches its optimal value.
First, it is necessary to make $\partial R/\partial \mu^{a}=0$ and $\partial R/\partial \mu^{b}=0$. 
The corresponding results are 
\begin{equation}
    \begin{split}
        &\delta(\mu^{a}_{m})^{2}+\mu^{a}_{m}\mu^{b}_{m}-\mu^{b}_{m}=0,\\&(\mu^{b}_{m})^{2}+\delta\mu^{a}_{m}\mu^{b}_{m}-\delta\mu^{a}_{m}=0.
    \end{split}
\end{equation}
Then, a straightforward calculation reveals that if $\delta=1$, $\mu^{a}_{m}=\mu^{b}_{m}=0.5$; otherwise, if $\delta>1$, $\mu^{a}_{m}=\frac{\sqrt{\delta}-1}{\delta-1}$ and $\mu^{b}_{m}=\frac{\delta-\sqrt{\delta}}{\delta-1}$ (where negative values are rounded off).
Finally, the above results can be summarized as
\begin{equation}
    \begin{split}
        &\mu^{a}_{m}+\mu^{b}_{m}=1, \\&\frac{\mu^{b}_{m}}{\mu^{a}_{m}}=\sqrt{\delta}=\sqrt{\frac{\eta^{a}}{\eta^{b}}}.
    \end{split}
\end{equation}

When taking into account the dark count rate and the approximation error of the Taylor series, the equalities in the above equations become the approximations, as described in Eq. (\ref{lam0}).

\subsection{Proof of Equation (\ref{lam1})}\label{appB2}
The maximal pairing interval is $\lambda=1$ in Eq. (\ref{lam1}).
The approximations in the formula are influenced not just by the above errors but also by the probability of successful detection $p$. 
Note that the effect of $p$ on the optimal intensities is negligible.
We assume that $p\ll 1$ in the following derivation.

The corresponding pairing ratio $r_{p}(p,\lambda)$ is 
\begin{equation}
    \begin{split}
        r_{p}(p,\lambda)=\frac{p^{2}}{1+p}\approx p^{2},
    \end{split}
\end{equation}
where the approximation is due to $p\ll 1$.

Therefore, the key-rate formula for this case is
\begin{equation}
    \begin{split}
        R&=r_{p}(p,\lambda) r_{s} \left \{ \bar{q}_{(1,1)} \left [ 1-H(e_{(1,1)})\right ]-fH(e^{(\mu^{a},\mu^{b}),Z}) \right \}\\&\approx\frac{1-H(4\%)}{8}\eta^{a}\eta^{b}\mu^{a}\mu^{b}e^{-\mu^{a}}e^{-\mu^{b}}.
    \end{split}
\end{equation}

Similarly, we set $\partial R/\partial \mu^{a}=0$ and $\partial R/\partial \mu^{b}=0$ to determine $\mu^{a}_{m}$ and $\mu^{b}_{m}$. 
The corresponding results are
\begin{equation}
    \begin{split}
        \mu^{a}_{m}\approx1,\quad \mu^{b}_{m}\approx1.
    \end{split}
\end{equation}
The approximations remain valid even when all influencing factors are considered, provided that $\Delta$ is relatively small.


\nocite{*}

\bibliography{apssamp}

\end{document}